\documentclass[a4paper,twoside,10pt,english,numbers=noenddot]{scrbook}

\usepackage[english]{babel}
\makeatletter
\def\month@ngerman{\ifcase\month \or Januar\or Februar\or M\"arz\or April\or Mai\or Juni\or Juli\or August\or September\or Oktober\or November\or Dezember\fi}
\def\month@english{\ifcase\month \or January\or February\or March\or April\or May\or June\or July\or August\or September\or October\or November\or December\fi}
\makeatother
\usepackage{hyphenat} 

\usepackage[T1]{fontenc}
\usepackage{helvet} 
\usepackage{indentfirst}
\usepackage[keeplastbox]{flushend} 

\usepackage[usenames,dvipsnames]{xcolor} 
\usepackage{graphicx} 
\usepackage{wrapfig} 
\usepackage{tikz}
\usepackage{rotating}

\usepackage{ifthen}
\usepackage{forloop}
\usepackage{etoolbox}

\usepackage[fleqn]{amsmath} 
\usepackage{amssymb} 
\usepackage{amsfonts} 
\usepackage{amstext} 
\usepackage{wasysym}
\usepackage{mathrsfs} 
\usepackage{mathtools}
\usepackage{upgreek}
\usepackage{siunitx}
\usepackage{esint} 
\usepackage{cancel} 
\usepackage{empheq}
\allowdisplaybreaks

\usepackage{multicol} 
\usepackage{multirow} 
\usepackage{dcolumn}
\usepackage{tabularx} 
\newcolumntype{L}[1]{>{\raggedright\arraybackslash\hsize=#1\hsize}X}
\newcolumntype{R}[1]{>{\raggedleft\arraybackslash\hsize=#1\hsize}X}
\newcolumntype{C}[1]{>{\centering\arraybackslash\hsize=#1\hsize}X}
\usepackage{dblfloatfix} 

\usepackage{enumitem} 
\setlist{nosep} 

\usepackage[a4paper,portrait]{geometry}
\newlength{\TwoColumnWidth}
\newlength{\OneColumnWidth}
\usepackage[headsepline]{scrlayer-scrpage}

\usepackage[hang,bottom]{footmisc}
\deffootnote{\footnotemargin}{0pt}{%
	\textsuperscript{\thefootnotemark}
}
\usepackage[singlelinecheck=false,font=small,labelfont=bf,format=plain]{caption} 
\usepackage[singlelinecheck=true,font=small,labelfont=bf,format=plain]{subfig}
\setkomafont{sectioning}{\rmfamily\bfseries}
\setkomafont{descriptionlabel}{\rmfamily\bfseries}

\usepackage[subfigure]{tocloft} 
\addto\captionsenglish{

}
\makeatletter
  \renewcommand*{\@pnumwidth}{20pt} 
  \renewcommand*{\@tocrmarg}{30pt plus 5pt minus 0pt} 
  \renewcommand*{\@dotsep}{4} 
\makeatother

\newcounter{chapterappendixcounter}[chapter]

\newcounter{totalpagecounter}
\newcounter{totalfigurecounter}
\newcounter{totaltablecounter}
\newcounter{totalcitecounter}
\newcounter{totalpages}
\newcounter{totalfigures}
\newcounter{totaltables}
\newcounter{totalcites}
\makeatletter
\@input{\jobname.myaux}
\makeatother

\definecolor{white}{rgb}{1,1,1}
\definecolor{black}{rgb}{0,0,0}
\definecolor{red}{rgb}{1,0,0}
\definecolor{green}{rgb}{0,1,0}
\definecolor{blue}{rgb}{0,0,1}
\definecolor{cyan}{rgb}{0,1,1}
\definecolor{magenta}{rgb}{1,0,1}
\definecolor{yellow}{rgb}{1,1,0}
\definecolor{darkgreen}{rgb}{0,0.6,0}
\definecolor{darkyellow}{rgb}{0.8,0.8,0}
\definecolor{orange}{rgb}{1,0.5,0}
\definecolor{tuc}{RGB}{0,90,70}
\definecolor{tuclight}{RGB}{218,234,194}
\definecolor{tucorange}{RGB}{242,148,0}
\definecolor{tucbg}{RGB}{224,233,233}

\usepackage[
	colorlinks=true,                   
	urlcolor=blue,                     
	filecolor=blue,                    
	linkcolor=blue,                    
	citecolor=blue,                    
	breaklinks=true,                   
	backref=page,                      
	pagebackref=true,                  
	bookmarks=true,                    
	bookmarksopen=false,               
	bookmarksnumbered=true,            
	pdfauthor={Fabian Teichert},       
	pdfdisplaydoctitle=true,           
	pdfstartview=FitH,                 
	pdfpagelayout=TwoPageRight,        
]{hyperref}

\newif\ifsinglepaper\singlepaperfalse
\addto\captionsenglish{
	\renewcommand\bibname{References}
}

\renewcommand*{\backref}[1]{}
\renewcommand*{\backrefalt}[4]{%
	\ifsinglepaper\else%
		\ifcase #1
		\or (cited at p.~#2).
		\else (cited at pp.~#2).
		\fi%
	\fi
}
\usepackage{cite} 

\newcommand{\bstindent}{99}
\newcommand{\bstaddress}{}
\newcommand{\bstauthor}{}

\newcommand{\bstjournal}{}

\newcommand{\bstpublisher}{}
\newcommand{\bstinstitution}{}
\newcommand{\bsttitle}{\itshape}
\newcommand{\bstvolume}{}
\newcommand{\bstyear}{}
\newcommand{\bbland}{and}

\newcommand{\bblnov}{November}

\newcommand\bibliographysection{\section}
\newcommand\bibliographysectionstyle{}
\newcommand\bibliographyitemsize{\normalsize}
\newcommand\bibliographyitemseparation{}
\makeatletter
\newcommand\bibcontentsline{\addcontentsline{toc}{section}{References}}
\renewenvironment{thebibliography}[1]{%
	\bibliographysection{\bibliographysectionstyle\bibname}
	\bibcontentsline%
	\renewcommand\rightmark{\bibname}
	\list{\@biblabel{\@arabic\c@enumiv}}{\settowidth\labelwidth{\@biblabel{#1}}%
		\leftmargin\labelwidth%
		\advance\leftmargin\labelsep%
		\@openbib@code%
		\usecounter{enumiv}%
		\let\p@enumiv\@empty%
		\renewcommand\theenumiv{\@arabic\c@enumiv}%
	}%
	\sloppy
	\clubpenalty4000
	\@clubpenalty \clubpenalty
	\widowpenalty4000%
	\sfcode`\.\@m%
}{%
	\def\@noitemerr{\@latex@warning{Empty `thebibliography' environment}}%
	\endlist%
}
\makeatother
\let\oldthebibliography\thebibliography
\renewcommand\thebibliography[1]{
	\bibliographyitemsize
	\oldthebibliography{#1}
	\bibliographyitemseparation
}

\let\oldtwocolumn\twocolumn
\let\oldonecolumn\onecolumn
\newif\iftwocolumn\twocolumntrue
\def\onecolumn{\twocolumnfalse}
\def\twocolumn{\twocolumntrue}
\newif\ifarticlestyle\articlestylefalse

\renewcommand{\title}[2]{%
	\setcounter{authors}{0}%
	\setcounter{addresses}{0}%
	\setcounter{keywords}{0}%
	\def\inserttitle{#1}%
	\def\articlelabel{#2}
}
\newcommand{\email}[1]{\def\insertemail{#1}}
\newcommand{\abstract}[1]{\def\insertabstract{#1}}
\def\insertjournal{}
\def\insertjournalshort{}
\def\insertdoi{}
\def\insertarxiv{}
\def\insertarxivshort{}
\newcommand{\journal}[4][nothing]{%
	\def\insertjournalshort{#2}%
	\if\relax\detokenize{#3}\relax%
		\def\insertjournal{}%
	\else%
		\def\tmpa{#1}%
		\def\tmpb{submitted}%
		\def\tmpc{accepted}%
		\ifx\tmpa\tmpb%
			\def\journalpre{Submitted to: }%
		\else%
			\ifx\tmpa\tmpc%
				\def\journalpre{Accepted in: }%
			\else%
				\def\journalpre{}%
			\fi%
		\fi%
		\if\relax\detokenize{#4}\relax\def\insertjournal{\journalpre#3}\else\def\insertjournal{\journalpre\href{#4}{#3}}\fi%
	\fi%
}
\newcommand{\doi}[1]{\if\relax\detokenize{#1}\relax\def\insertdoi{}\else\def\insertdoi{DOI: \href{http://dx.doi.org/#1}{#1}}\fi}
\newcommand{\arxiv}[2]{\if\relax\detokenize{#1}\relax\def\insertarxiv{}\def\insertarxivshort{}\else\def\insertarxiv{arXiv: \href{https://arxiv.org/abs/#1}{#1 [#2]}}\def\insertarxivshort{arXiv: #1}\fi}
\newcounter{authors}
\newcounter{addresses}
\newcounter{keywords}
\newcommand{\addauthor}[2]{\csdef{author\arabic{authors}}{#1}\csdef{authoraddress\arabic{authors}}{#2}\stepcounter{authors}}
\newcommand{\addaddress}[1]{\csdef{address\arabic{addresses}}{#1}\stepcounter{addresses}}
\newcommand{\addkeyword}[1]{\csdef{keyword\arabic{keywords}}{#1}\stepcounter{keywords}}

\newcounter{otherchapter}
\newcounter{normalchapter}
\newcounter{othercounter}
\newcounter{i}
\newcounter{j}
\makeatletter
\let\normalchapter\chapter

\renewcommand\chapter{%
	\@ifstar{%
		\normalchapter*%
	}{%
		\stepcounter{normalchapter}%
		\normalchapter%
	}%
}
\newcommand\otherchapter{%
	\protected@write\@auxout{}{\string\@writefile{lof}{\string\addvspace{10\string\p@}}}%
	\protected@write\@auxout{}{\string\@writefile{lot}{\string\addvspace{10\string\p@}}}%
	\scr@startsection{chapter}{1}{\z@}{0ex \@plus -0.2ex}{3.5ex \@plus 0.2ex}{\Large\bfseries}%
}
\newcommand\reftype{}
\newcommand\reflabel{}
\newcommand\refnumber{}
\newcommand\refshortnumber{}
\newcommand\reftext{}
\newcommand{\articletitlesub}{%
	\renewcommand\reftype{chapter}%
	\renewcommand\reflabel{section*.\arabic{othercounter}}%
	\renewcommand\refnumber{\Alph{otherchapter}}%
	\renewcommand\refshortnumber{}%
	\renewcommand\reftext{\inserttitle\ifx\insertjournalshort\empty\ (\insertarxivshort)\else\ (\insertjournalshort)\fi}%
	\pdfbookmark[0]{\reftext}{\reflabel}%
	\otherchapter*{\inserttitle}%
	\protected@write\@auxout{}{\string\@writefile{toc}{\string\contentsline {\reftype}{\string\numberline {\refnumber}\reftext}{\thepage}{\reflabel}}}%
	\Alabel{\articlelabel}%
	\noindent\textbf{%
		\large\csuse{author0}$^{\csuse{authoraddress0}}$%
		\forloop{i}{1}{\value{i} < \value{authors}}{%
			, \csuse{author\arabic{i}}$^{\csuse{authoraddress\arabic{i}}}$%
		}
	}\\[1em]
	\normalsize
	\setcounter{j}{0}
	\forloop{i}{0}{\value{i} < \value{addresses}}{%
		\stepcounter{j}
		$^{\arabic{j}}$\,\csuse{address\arabic{i}}
		\ifthenelse{\value{j}<\value{addresses}}{\\}{}
	}
	\ifx\insertemail\empty\\[1em]\else\\[0.5em]E-mail address: \insertemail\\[1em]\fi
	\textbf{Abstract:} \insertabstract
	\ifthenelse{\value{keywords}=0}{}{
		\\[1em]
		Keywords: \csuse{keyword0}%
			\forloop{i}{1}{\value{i} < \value{keywords}}{%
				; \csuse{keyword\arabic{i}}%
			}
	}
}
\newcommand{\articletitle}{%
	\setcounter{articlepage}{0}%
	\stepcounter{otherchapter}%
	\stepcounter{chapter}%
	\setcounter{section}{0}%
	\setcounter{subsection}{0}%
	\setcounter{subsubsection}{0}%
	\stepcounter{othercounter}%
	\iftwocolumn\oldtwocolumn[\articletitlesub\vspace{1.5em}]\else\oldonecolumn\articletitlesub\fi%
}%
\let\oldchapter\chapter
\let\oldsection\section
\let\oldsubsection\subsection
\let\oldsubsubsection\subsubsection
\newcommand\articlesectiondata[1]{%
	\renewcommand\reftype{section}%
	\renewcommand\reflabel{section.\arabic{chapter}.\arabic{section}}%
	\renewcommand\refnumber{\Alph{otherchapter}.\arabic{section}}%
	\renewcommand\refshortnumber{\arabic{section}}%
	\renewcommand\reftext{#1}%
}
\newcommand\articlesubsectiondata[1]{%
	\renewcommand\reftype{subsection}%
	\renewcommand\reflabel{subsection.\arabic{chapter}.\arabic{section}.\arabic{subsection}}%
	\renewcommand\refnumber{\Alph{otherchapter}.\arabic{section}.\arabic{subsection}}%
	\renewcommand\refshortnumber{\arabic{section}.\arabic{subsection}}%
	\renewcommand\reftext{#1}%
}
\newcommand\articlesectionnostar[1]{%
	\articlesectiondata{#1}%
	\pdfbookmark[1]{\reftext}{\reflabel}%
	\scr@startsection{section}{1}{\z@}{-3.5ex \@plus -1ex \@minus -0.2ex}{2.3ex \@plus 0.2ex}{\normalfont\bfseries}{#1}%
	\protected@write\@auxout{}{\string\@writefile{toc}{\string\contentsline {\reftype}{\string\numberline {\refnumber}#1}{\thepage}{\reflabel}}}%
}
\newcommand\articlesectionstar[1]{%
	\articlesectiondata{#1}%
	\scr@startsection{section}{1}{\z@}{-3.5ex \@plus -1ex \@minus -0.2ex}{2.3ex \@plus 0.2ex}{\normalfont\bfseries}*{#1}%
}
\newcommand\articlesubsectionnostar[1]{%
	\articlesubsectiondata{#1}%
	\pdfbookmark[2]{\reftext}{\reflabel}%
	\scr@startsection{subsection}{2}{\z@}{-3.5ex \@plus -1ex \@minus -0.2ex}{2.3ex \@plus 0.2ex}{\normalfont\bfseries}{#1}%
	\protected@write\@auxout{}{\string\@writefile{toc}{\string\contentsline {\reftype}{\string\numberline {\refnumber}#1}{\thepage}{\reflabel}}}%
}
\newcommand\articlesubsectionstar[1]{%
	\articlesubsectiondata{#1}%
	\scr@startsection{subsection}{2}{\z@}{-3.5ex \@plus -1ex \@minus -0.2ex}{2.3ex \@plus 0.2ex}{\normalfont\bfseries}*{#1}%
}
\newcommand\articlesection{\@ifstar{\stepcounter{othercounter}\articlesectionstar}{\articlesectionnostar}}
\newcommand\articlesubsection{\@ifstar{\stepcounter{othercounter}\articlesubsectionstar}{\articlesubsectionnostar}}
\renewcommand\chapter{\@ifstar{\stepcounter{othercounter}\oldchapter*}{\oldchapter}}
\renewcommand\section{\@ifstar{\stepcounter{othercounter}\oldsection*}{\oldsection}}
\renewcommand\subsection{\@ifstar{\stepcounter{othercounter}\oldsubsection*}{\oldsubsection}}
\renewcommand\subsubsection{\@ifstar{\stepcounter{othercounter}\oldsubsubsection*}{\oldsubsubsection}}
\newcommand\listof{}

\newcommand\articlefiguredata{%
	\renewcommand\listof{lof}%
	\renewcommand\reftype{figure}%
	\renewcommand\reflabel{figure.\arabic{chapter}.\arabic{figure}}%
	\renewcommand\refnumber{\Alph{otherchapter}.\arabic{figure}}%
	\renewcommand\refshortnumber{\arabic{figure}}%
}
\newcommand\articletabledata{%
	\renewcommand\listof{lot}%
	\renewcommand\reftype{table}%
	\renewcommand\reflabel{table.\arabic{chapter}.\arabic{table}}%
	\renewcommand\refnumber{\Alph{otherchapter}.\arabic{table}}%
	\renewcommand\refshortnumber{\arabic{table}}%
}
\renewenvironment{figure}{\articlefiguredata\begin{oldfigure}}{\end{oldfigure}} 
\renewenvironment{table}{\articletabledata\begin{oldtable}}{\end{oldtable}} 
\newenvironment{articlefigure}{\articlefiguredata\begin{figure}}{\end{figure}}
\newenvironment{articlefigure*}{\articlefiguredata\begin{figure*}}{\end{figure*}}
\newenvironment{articletable}{\articletabledata\begin{table}}{\end{table}}
\newenvironment{articletable*}{\articletabledata\begin{table*}}{\end{table*}}
\let\oldcaption\caption
\newcommand\Acaption[2][]{%
	\oldcaption[#1]{#2}%
	\renewcommand\reftext{#1}%
	\protected@write\@auxout{}{\string\@writefile{\listof}{\string\contentsline {\reftype}{\string\numberline {\refnumber}#1}{\thepage}{\reflabel}}}%
}
\renewcommand\caption[2][]{\ifarticlestyle\Acaption[#1]{#2}\else\oldcaption[#1]{#2}\fi}

\newcommand\botholdlabel[1]{\oldlabel{#1}\oldlabel{A#1}}
\newenvironment{articleequation}{\begin{equation}\renewcommand\label{\botholdlabel}}{\end{equation}} 
\newcommand\Aref[1]{\oldref{A#1}}
\newcommand\Alabel[1]{%
	\protected@write\@auxout{}{\string\newlabel{#1}{{\refnumber}{\thepage}{\reftext}{\reflabel}{}}}%
	\protected@write\@auxout{}{\string\newlabel{A#1}{{\refshortnumber}{\thepage}{\reftext}{\reflabel}{}}}%
}
\AtBeginDocument{%
	\let\oldref\ref%
	\let\oldlabel\label%
	\renewcommand\ref[1]{\ifarticlestyle\Aref{#1}\else\oldref{#1}\fi}%
	\renewcommand\label[1]{\ifarticlestyle\Alabel{#1}\else\oldlabel{#1}\fi}%
}
\makeatother

\newcommand\articlestyleheaderleft{%
	\ifx\insertarxiv\empty%
		\ifx\insertjournal\empty\linebreak\textnormal\insertdoi\else\linebreak\textnormal\insertjournal\fi%
	\else%
		\ifx\insertdoi\empty\linebreak\textnormal\insertjournal\else\textnormal\insertjournal\linebreak\textnormal\insertdoi\fi%
	\fi%
}
\newcommand\articlestyleheaderright{%
	\ifx\insertarxiv\empty%
		\ifx\insertjournal\empty\else\linebreak\textnormal\insertdoi\fi%
	\else%
		\linebreak\textnormal\insertarxiv%
	\fi%
}
\newcommand\articlestyleheadercenter{%
	\linebreak\textnormal\thechapter
}

\newcounter{articlepage}
\newcommand\articlepagemark{\arabic{articlepage}}

\newcommand\nocontentsline[3]{}
\let\oldaddcontentsline\addcontentsline
\newcommand\normalstyle{%
	\articlestylefalse%
	\KOMAoptions{fontsize=11pt}%
	\newgeometry{left=3cm,right=2.5cm,top=4cm,bottom=4cm}
	\setlength{\headheight}{26pt}
	\setlength{\headsep}{24pt}
	\setlength{\footskip}{30pt}
	\setlength{\TwoColumnWidth}{\textwidth}
	\setlength{\OneColumnWidth}{0.5\TwoColumnWidth-0.5\columnsep}
	\clearpairofpagestyles%
	\ihead{}%
	\chead{}%
	\ohead{\ifthispageodd{\textnormal\rightmark}{\textnormal\leftmark}}%
	\ifoot{}%
	\cfoot{}%
	\ofoot[\textnormal\pagemark]{\textnormal\pagemark}%
	\let\addcontentsline\oldaddcontentsline%
	\renewcommand{\thechapter}{\arabic{normalchapter}}%
	\renewcommand{\thesection}{\arabic{normalchapter}.\arabic{section}}%
	\renewcommand{\thesubsection}{\arabic{normalchapter}.\arabic{section}.\arabic{subsection}}%
	\renewcommand{\thesubsubsection}{\arabic{normalchapter}.\arabic{section}.\arabic{subsection}.\arabic{subsubsection}}%
	\renewcommand{\thefigure}{\arabic{normalchapter}.\arabic{figure}}%
	\renewcommand{\thetable}{\arabic{normalchapter}.\arabic{table}}%
	\renewcommand{\theequation}{\arabic{normalchapter}.\arabic{equation}}%
	\renewcommand\bibliographysection{\section*}%
	\renewcommand\bibcontentsline{\addcontentsline{toc}{section}{References}}
	\renewcommand\bibliographysectionstyle{}%
	\renewcommand\bibliographyitemsize{\normalsize}%
	\renewcommand\bibliographyitemseparation{%
		\setlength{\parskip}{0pt}%
		\setlength{\itemsep}{5pt plus 0.3ex}%
	}%
	\allowdisplaybreaks%
}
\newcommand\articlestyle{%
	\articlestyletrue%
	\KOMAoptions{fontsize=10pt}%
	\newgeometry{left=1.5cm,right=1.5cm,top=2.95cm,bottom=1.55cm}
	\setlength{\headheight}{24pt}
	\setlength{\headsep}{20pt}
	\setlength{\footskip}{1.2cm}
	\setlength{\TwoColumnWidth}{\textwidth}
	\setlength{\OneColumnWidth}{0.5\TwoColumnWidth-0.5\columnsep}
	\clearpairofpagestyles%
	\ihead{\ifthispageodd{\articlestyleheaderleft}{\articlestyleheaderright}}%
	\chead{\ifsinglepaper\else\articlestyleheadercenter\fi}%
	\ohead{\ifthispageodd{\articlestyleheaderright}{\articlestyleheaderleft}}%
	\ifoot{}%
	\cfoot{\ifsinglepaper\textnormal\pagemark\else\stepcounter{articlepage}\textnormal{\thechapter-\articlepagemark}\fi}%
	\ofoot{\ifsinglepaper\else\textnormal\pagemark\fi}%
	\let\addcontentsline\nocontentsline%
	\renewcommand{\thechapter}{\Alph{otherchapter}}%
	\renewcommand{\thesection}{\arabic{section}}%
	\renewcommand{\thesubsection}{\arabic{section}.\arabic{subsection}}%
	\renewcommand{\thesubsubsection}{\arabic{section}.\arabic{subsection}.\arabic{subsubsection}}%
	\renewcommand{\thefigure}{\arabic{figure}}%
	\renewcommand{\thetable}{\arabic{table}}%
	\renewcommand{\theequation}{\arabic{equation}}%
	\renewcommand\bibliographysection{\articlesection*}%
	\renewcommand\bibcontentsline{\oldaddcontentsline{toc}{section}{References}}
	\renewcommand\bibliographysectionstyle{\normalsize}%
	\renewcommand\bibliographyitemsize{\small}%
	\renewcommand\bibliographyitemseparation{%
		\setlength{\parskip}{0pt}%
		\setlength{\itemsep}{0pt plus 0.3ex}%
	}%
	\interdisplaylinepenalty=10000%
}
\normalstyle

\renewcommand{\hbar}{\mathchar'26\mkern-9mu \mathrm{h}}

\newcommand{\hamilton}{\mathcal{H}}
\newcommand{\overlap}{\mathcal{O}}
\newcommand{\coupling}{\tau}

\newcommand{\green}{\mathcal{G}}

\newcommand{\transmission}{\mathcal{T}}

\newcommand{\imag}{\text{i}}

\singlepapertrue
\begin{document}

\raggedbottom

\frontmatter
\clearpairofpagestyles
\ofoot[\textnormal\pagemark]{\textnormal\pagemark}
\KOMAoptions{headsepline=false}

\mainmatter
\KOMAoptions{headsepline=true}

\normalstyle
\articlestyle
\normalsize

\renewcommand{\hamilton}{\mathcal{H}}
\renewcommand{\coupling}{\tau}
\renewcommand{\overlap}{\mathcal{S}}
\renewcommand{\green}{\mathcal{G}}
\renewcommand{\transmission}{\mathcal{T}}
\renewcommand{\imag}{\text{i}}

\twocolumn 

\title{Electronic transport in metallic carbon nanotubes with mixed defects within the strong localization regime}{CMS}

\addauthor{Fabian Teichert}{1,3,4}
\addauthor{Andreas Zienert}{2}
\addauthor{J\"org Schuster}{3,4}
\addauthor{Michael Schreiber}{1,4}

\addaddress{Institute of Physics, Chemnitz University of Technology, 09107 Chemnitz, Germany}
\addaddress{Center for Microtechnologies, Chemnitz University of Technology, 09107 Chemnitz, Germany}
\addaddress{Fraunhofer Institute for Electronic Nano Systems (ENAS), 09126 Chemnitz, Germany}
\addaddress{Dresden Center for Computational Materials Science (DCMS), TU Dresden, 01062 Dresden, Germany}

\email{fabian.teichert@physik.tu-chemnitz.de}

\abstract{
We study the electron transport in metallic carbon nanotubes (CNTs) with realistic defects of different types.
We focus on large CNTs with many defects in the mesoscopic range.
In a recent paper we demonstrated that the electronic transport in those defective CNTs is in the regime of strong localization.
We verify by quantum transport simulations that the localization length of CNTs with defects of mixed types can be related to the localization lengths of CNTs with identical defects by taking the weighted harmonic average.
Secondly, we show how to use this result to estimate the conductance of arbitrary defective CNTs, avoiding time consuming transport calculations.
}

\addkeyword{carbon nanotube (CNT)}
\addkeyword{defect}
\addkeyword{density-functional-based tight binding (DFTB)}
\addkeyword{electronic transport}
\addkeyword{recursive Green's function formalism (RGF)}
\addkeyword{strong localization}

\journal{Comput. Mater. Sci. 138 (2017), 49--57}{Computational Materials Science 138 (2017), 49--57}{http://www.sciencedirect.com/science/article/pii/S0927025617302987} 
\doi{10.1016/j.commatsci.2017.06.001} 
\arxiv{1705.01749}{cond-mat.mes-hall} 

\articletitle

\articlesection{Introduction}

Carbon nanotubes (CNTs) offer a large variety of properties~\cite{RevModPhys.79.677, NanoRes.1.361, JPhysCondMat.24.313202}, which can be very useful for future electronic devices.
One of them is the very high conductance in the ballistic regime~\cite{Nature.393.240} that makes CNTs attractive for metallic interconnect systems~\cite{MicroEng.64.399, MicroEng.120.188, MicroEng.120.210, Dissertation.Fiedler}.
Although research on CNTs has continued for many years since their discovery in 1991 and clean CNTs approaching the theoretical conductance limit can be produced under well-defined laboratory conditions \cite{PhysRevLett.87.106801}, current CNT-based devices at the wafer level which means a fast and reproducible fabrication are still not reaching that limit.
One reason is the strong impact of defects~\cite{Science.272.523, Nature.382.54}, which cannot be avoided during production processes at the wafer level~\cite{PhysRevB.63.245405, NatureMaterials.4.534, ComputMatSci.93.15, NanoLett.9.2285, JPhysDApplPhys.43.305402}, whether physically introduced like vacancies or chemically initiated like functionalizations.
They can, e.g., be caused by ion collisions within a gas atmosphere, by electron beam treatments, or within organic solutions, which are necessary steps for the fabrication of devices with difficult three dimensional geometries.
Thus, understanding the influence of these defects on electronic transport properties of CNTs is a necessary step towards their integration into microelectronic devices.

The present work approaches this subject at the theoretical level.
On the one hand, the size of such mesoscopic systems is of the order of hundred thousand atoms, and on the other hand, a statistical description with large ensembles has to be considered.
This is very time-consuming despite the availability of high performance computer resources and good scaling low-level methods.

In the past, most theoretical work in the field of quantum transport simulations focused on the properties of single selected CNTs, like it was done for vacancies~\cite{PhysRevLett.95.266801, JPhysCondMat.20.294214, JPhysCondMat.20.304211, JPhysCondMat.26.045303, JPhysChemC.116.1179}, substitutional atoms~\cite{SolidStateCommun.149.874, PhysRevLett.100.176803} and functionalizations~\cite{JPhysChemC.117.15266, PhysStatSolB.247.2962, NanoRes.3.288, NanoLett.9.940, PhysRevLett.100.176803}.
For this purpose, different electronic structure methods were addressed, from tight binding (e.g.~\cite{SolidStateCommun.149.874}) to density functional theory (e.g.~\cite{JPhysCondMat.20.294214, PhysRevLett.100.176803}).
These investigations showed that the conductance depends exponentially on the CNT length, what was also verified by experiments~\cite{NatureMaterials.4.534}.
This is an indication of the strong localization regime, which is also present in case of random Anderson disorder~\cite{PhysRevB.58.4882, PhysRevB.64.045409}.
But this regime does not only exist in CNTs.
Also other materials can exhibit Anderson disorder, e.g. silicon nanowires~\cite{PhysRevLett.99.076803}, showing that strong localization is an interesting and important transport regime in quasi-one-dimensional structures.

\begin{articlefigure*}
	\includegraphics{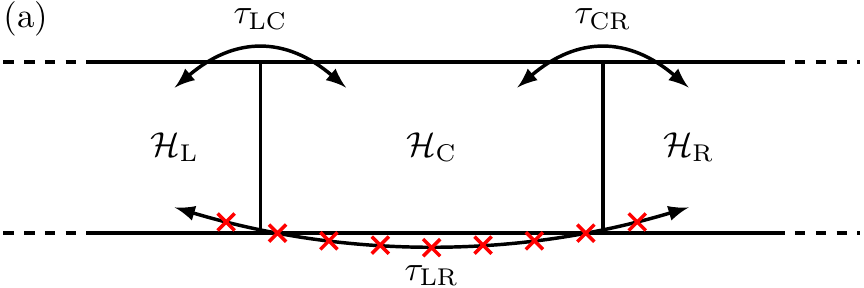}\hfill
	\includegraphics{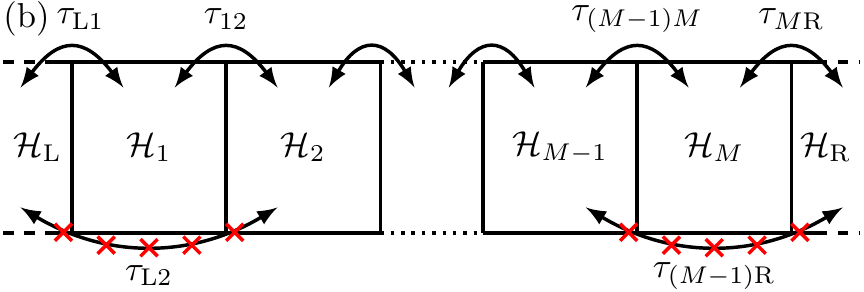}
	\caption[Scheme of a device system]{(Color online.) (a) The CNT as a device is divided into a (finite) central region C and two (half infinite) electrodes L and R. $\hamilton$ and $\coupling$ denote the Hamiltonian matrices and coupling matrices of the respective parts. (b) The central region is additionally divided into $M$ subsystems~\cite{NJPhys.16.123026}.}
	\label{CMS:fig:Device}
\end{articlefigure*}

Beyond the properties of single selected CNTs, a further description and quantification of the strong localization regime for different CNTs is necessary.
Flores began a systematic study by calculating the localization length for three metallic armchair CNTs~\cite{JPhysCondMat.20.304211}.
In a previous investigation~\cite{NJPhys.16.123026}, we continued these calculations for more CNTs to determine the diameter dependence.
Therein, we discussed CNTs with one type of defect.
In the following, we extend this work by calculating electronic transport properties of defective CNTs with defects of different types within one CNT.
Bringing all results together, we develop and explain a model for estimating the conductance of metallic CNTs with arbitrary diameter and an arbitrary number of different types of defects.

\articlesection{Theoretical framework}\label{CMS:Theory}

Electronic transport through mesoscopic systems can be described by quantum transport theory, which is done here in the equilibrium limit~\cite{Datta2005}.
The conductance formula in the limit of a small bias was introduced by Landauer and B\"uttiker~\cite{PhysRevB.31.6207}:
\begin{articleequation}
	G = -\text{G}_0\int\limits_{-\infty}^\infty\transmission(E)\frac{\text{d}f(E)}{\text{d}E}\,\text{d}E \quad .\label{CMS:eqn:Conductance}
\end{articleequation}%
$\text{G}_0 = 2\text{e}^2/\text{h}$ is the conductance quantum, $\transmission(E)$ the transmission function, and $f(E)$ the Fermi distribution, where the effect of temperature is included.

The transmission function can be calculated via the Schr\"odinger equation in a matrix representation and its Green's function.
For this purpose, the whole (infinite) CNT is treated as a device.
That means, it is divided into three main parts: the (finite) central region C, containing all the defects, and two (half infinite) electrodes L and R, as shown in figure \ref{CMS:fig:Device}(a).
Each part is described by Hamiltonian matrices $\hamilton_\text{L}$/$\hamilton_\text{C}$/$\hamilton_\text{R}$ and the coupling by similar matrices $\coupling_\text{LC}$/$\coupling_\text{CR}$.
The coupling $\coupling_\text{LR}$ can be neglected if the distance between the electrodes is large enough.
The electronic properties are calculated via the Green's function of the central region
\begin{articleequation}
	\green_\text{C} = \left[ (E + \imag\eta)\overlap - \hamilton_\text{C} - \varSigma_\text{L} - \varSigma_\text{R} \right]^{-1} \quad .\label{CMS:eqn:Green}
\end{articleequation}%
$\varSigma_\text{L} = \coupling_\text{CL}\green_\text{L}\coupling_\text{LC}$ and $\varSigma_\text{R} = \coupling_\text{CR} \green_\text{R} \coupling_\text{RC}$ are the self-energy matrices, which lead to an energetic shift due to the coupling to the electrodes.
The Green's functions of the electrodes $\green_\text{L/R}$ themselves can be calculated with the renormalization decimation algorithm (RDA), which is a fast iteration process~\cite{JPhysFMetPhys.14.1205, JPhysFMetPhys.15.851}.
$\eta$ is a small value for numerical stability, which shifts the singularities from the real axis into the complex plane\footnote{We use $\eta=10^{-7}$ for calculating $\green_\text{C}$ and $\eta=10^{-4}$ for calculating $\green_\text{L/R}$ via the RDA}.
$\overlap$ is the overlap matrix, which is present in cases where the representation is done in a non-orthogonal basis.
Altogether, this can be used to obtain the transmission function
\begin{articleequation}
	\transmission(E) = \text{Tr}\left(\varGamma_\text{R}\green_\text{C}\varGamma_\text{L}\green_\text{C}^\dagger\right) \quad .\label{CMS:eqn:Transmission}
\end{articleequation}%
$\varGamma_\text{L/R} = \imag(\varSigma_\text{L/R}-\varSigma_\text{L/R}^\dagger)$ are broadening matrices, which lead to an energetic broadening of each state due to the coupling to the electrodes.

In the following we want to treat CNTs of mesoscopic lengths with more than hundred thousand atoms in the defective central region, where the direct inversion (\ref{CMS:eqn:Green}) is too time-consuming.
Fortunately, when using a representation with localized basis functions, $\hamilton_\text{C}$ is block-tridiagonal.
The central region of our device can be subdivided into $M$~parts, where only neighbored parts are directly coupled, as shown in figure \ref{CMS:fig:Device}(b).
This simplifies (\ref{CMS:eqn:Transmission}) to
\begin{articleequation}
	\transmission(E) = \text{Tr}\left(\varGamma'_\text{R}\green_\text{M1}\varGamma'_\text{L}\green_\text{M1}^\dagger\right) \quad .\label{CMS:eqn:TransmissionRGF}
\end{articleequation}%
$\green_\text{M1}$ is the lower left block of $\green_\text{C}$. Its dimension is a factor $M$ smaller.
In the same way, $\varGamma_\text{L}'$ ($\varGamma_\text{R}'$) is the upper left (lower right) block of $\varGamma_\text{L}$ ($\varGamma_\text{R}$).
With the usage of the recursive Green's function formalism (RGF)~\cite{JPhysCSolidStatePhys.14.235}, $\green_\text{M1}$ can be calculated very efficiently within linearly scaling time $t=\mathcal{O}(M)$, which makes it possible at all to compute electronic transport properties of mesoscopic CNTs.
For this purpose, narrowing the cells in figure \ref{CMS:fig:Device}(b) lowers the computation time, which has to be taken into account when choosing these cells.

The following computations are performed neglecting phonon effects.
In the limit of a small bias, optical phonons have short coherence lengths of 180\,nm~\cite{NanoLett.4.517}, but also high energies above the thermal fluctuations.
They are not excitable.
Acoustic phonons have small energies of the order of thermal fluctuations and can be excited.
But their coherence length of 2400\,nm is much larger.
Therefore, inelastic scattering is not dominant for systems shorter than this length.
Beyond the following study, dephasing due to phonons can be included phenomenologically with the B\"uttiker probe model~\cite{PhysRevB.41.7411, JComputElectron.12.203}.
Here, the electron-phonon coupling strength is a parameter, which has to be assumed or calculated separately.
An additional self-consistency iteration cycle is necessary, raising the computation time.
E.g. the conductance of disordered graphene has been determined in this way~\cite{JComputElectron.12.203}.
Ab initio calculations of phonon modes and their influence on electron transport can also be done directly~\cite{PhysRevB.75.205413, PhysRevB.89.081405}, but are even more challenging.

\articlesection{Modeling details}

\begin{articlefigure}[tb]
	\newlength{\len}
	\setlength{\len}{0.11\textwidth}
	\begin{minipage}{0.21\len}~\end{minipage}\hfill
	\begin{minipage}{0.35\len}\centering UC\end{minipage}\hfill
	\begin{minipage}{0.35\len}\centering MV\end{minipage}\hfill
	\begin{minipage}{    \len}\centering MV$_\text{3H}$\end{minipage}\hfill
	\begin{minipage}{    \len}\centering DV$_\text{diag}$\end{minipage}\hfill
	\begin{minipage}{    \len}\centering DV$_\text{perp}$\end{minipage}\\
	\begin{minipage}{0.21\len}\begin{sideways}(5,5)-CNT\end{sideways}\end{minipage}\hfill
	\begin{minipage}{0.35\len}\includegraphics[width=\textwidth]{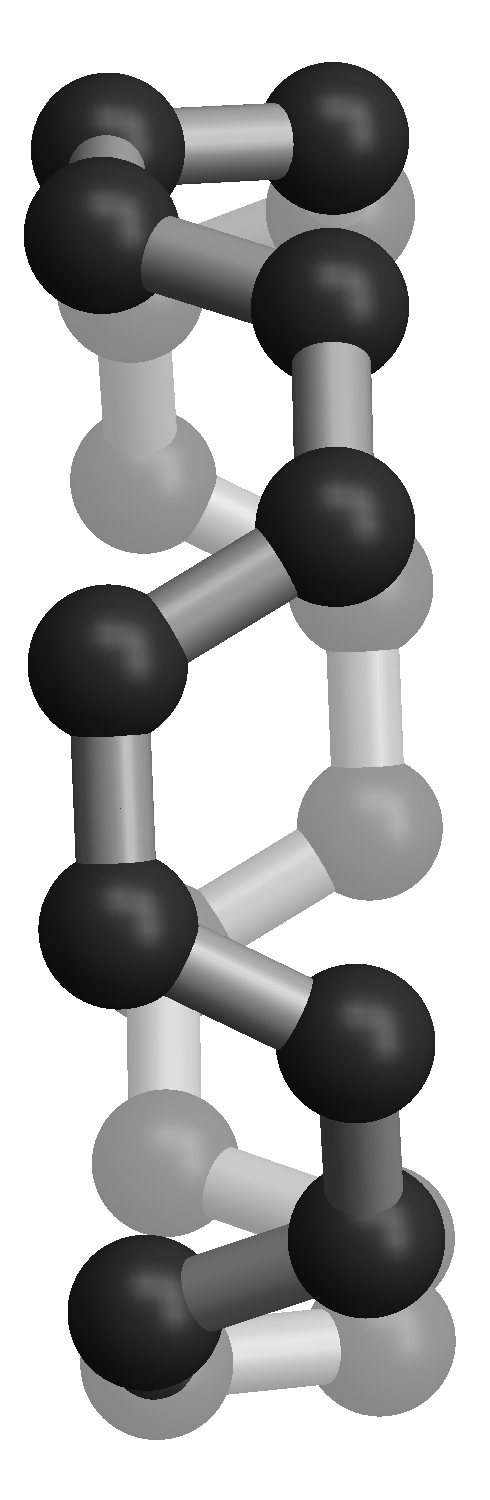}\end{minipage}\hfill
	\begin{minipage}{0.35\len}\includegraphics[width=\textwidth]{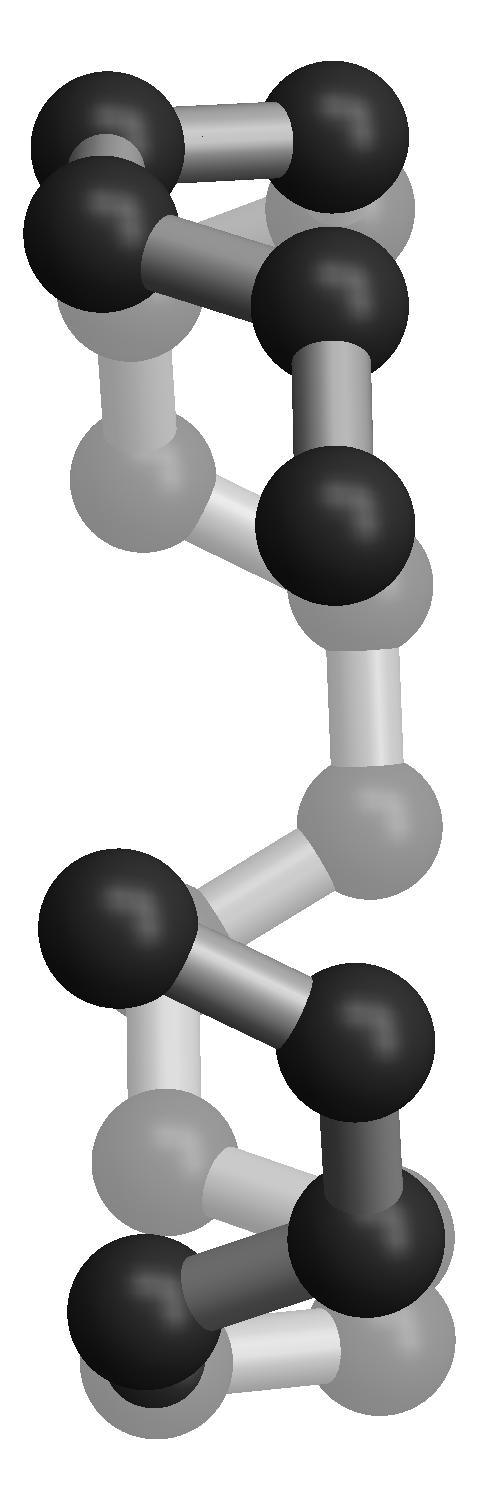}\end{minipage}\hfill
	\begin{minipage}{    \len}\includegraphics[width=\textwidth]{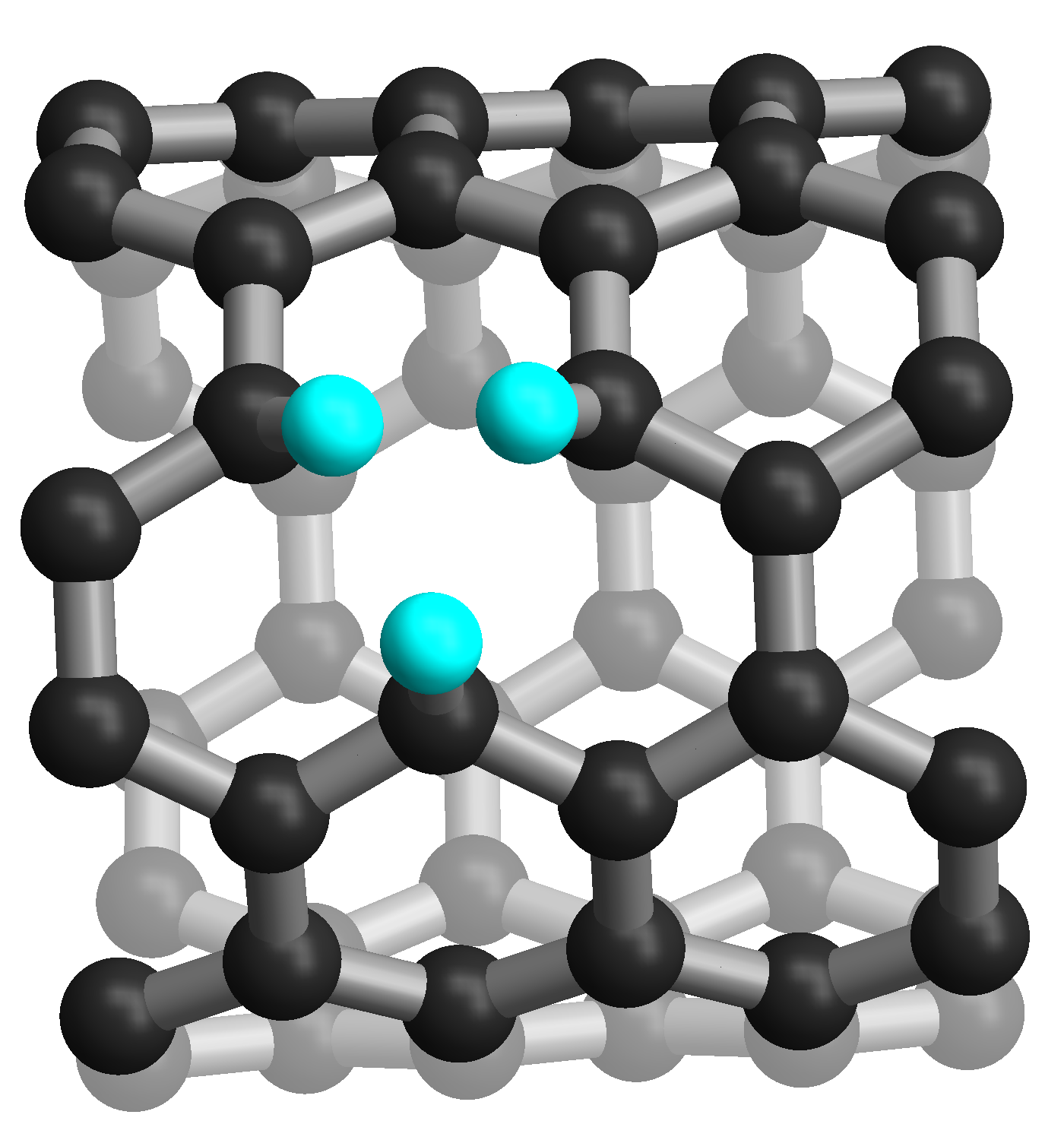}\end{minipage}\hfill
	\begin{minipage}{    \len}\includegraphics[width=\textwidth]{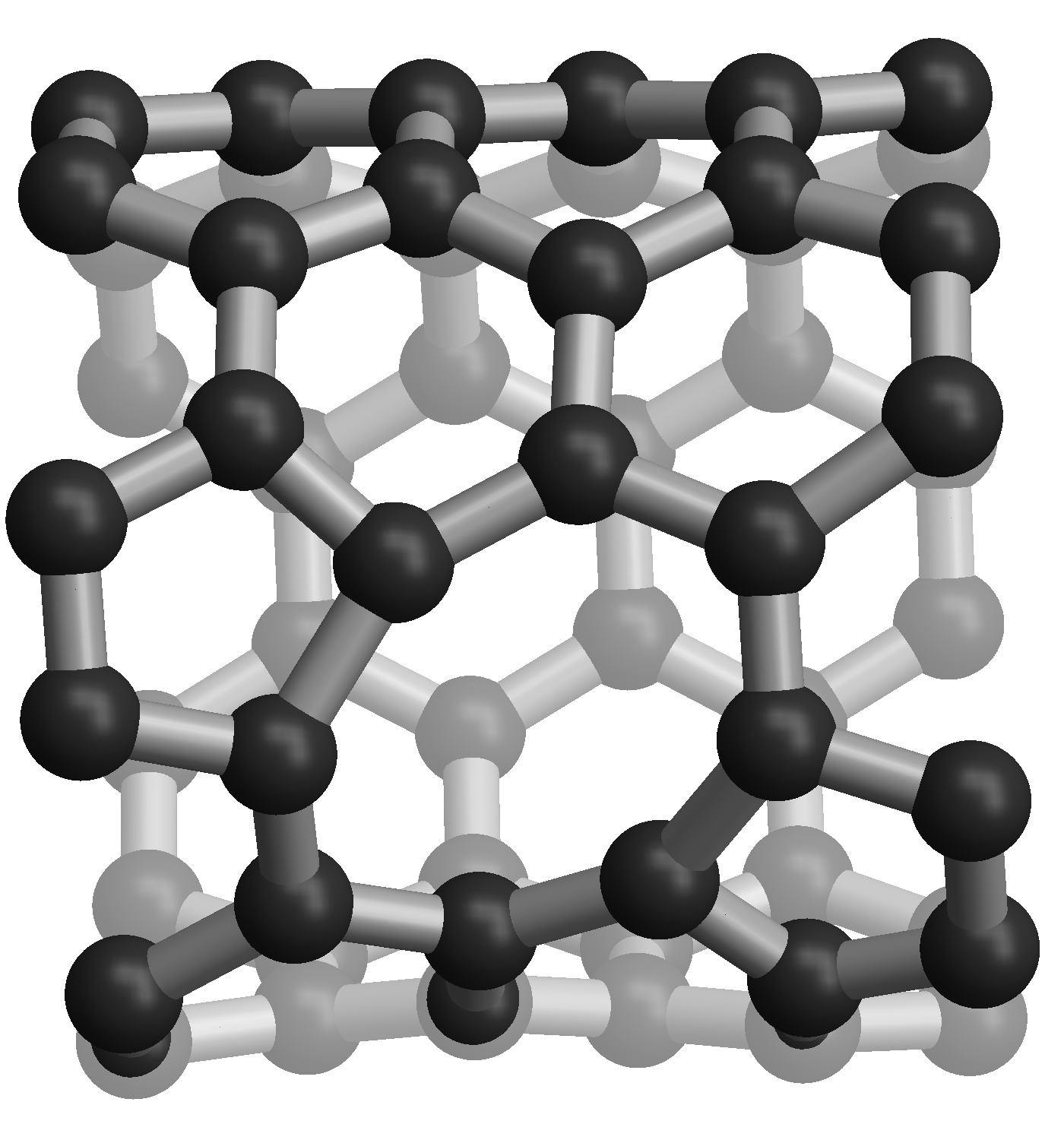}\end{minipage}\hfill
	\begin{minipage}{    \len}\includegraphics[width=\textwidth]{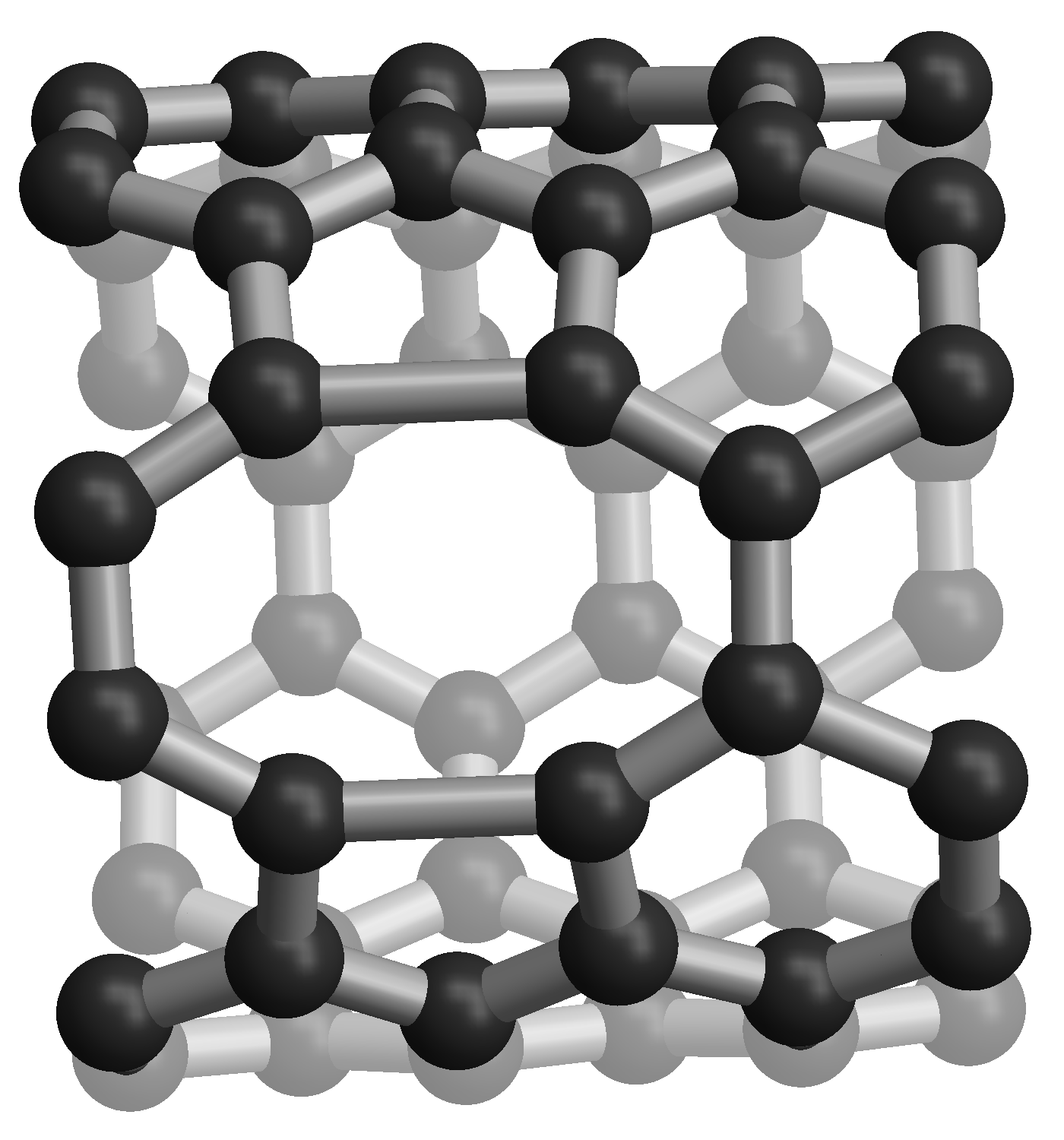}\end{minipage}\\
	\begin{minipage}{0.21\len}\begin{sideways}(10,10)-CNT\end{sideways}\end{minipage}\hfill
	\begin{minipage}{0.35\len}\includegraphics[width=\textwidth]{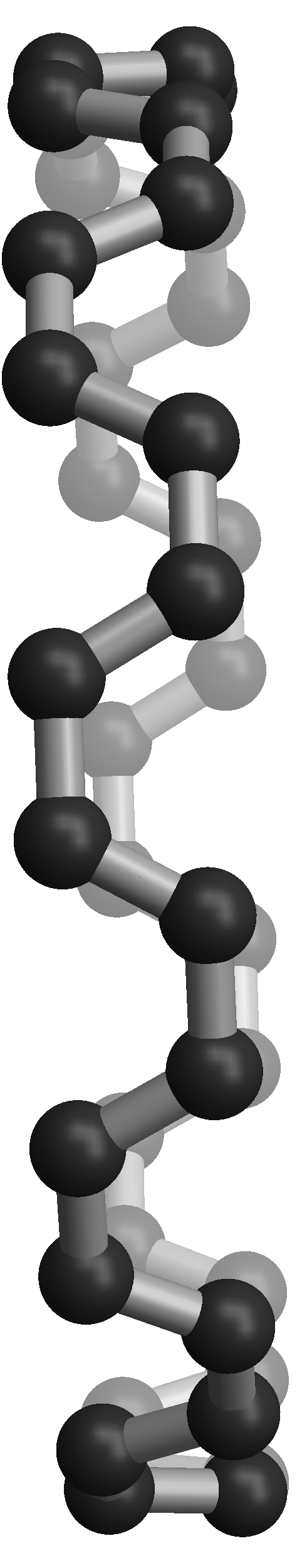}\end{minipage}\hfill
	\begin{minipage}{0.35\len}\includegraphics[width=\textwidth]{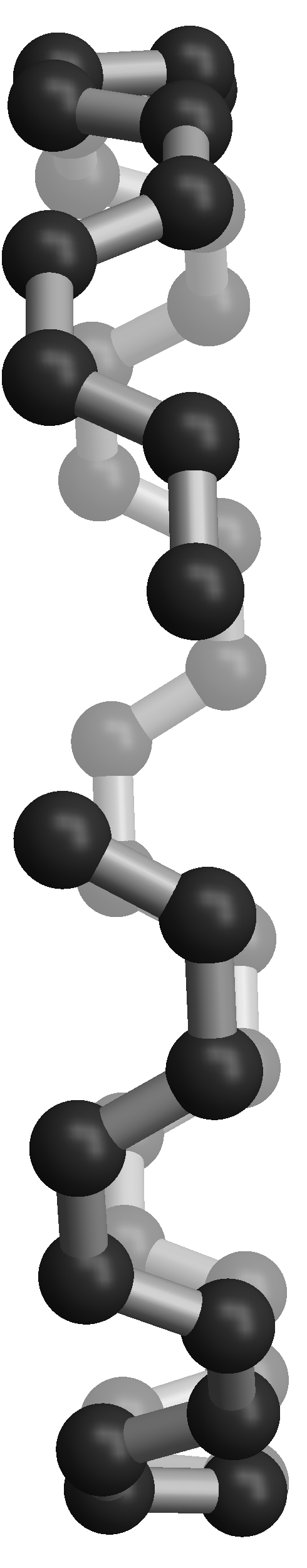}\end{minipage}\hfill
	\begin{minipage}{    \len}\includegraphics[width=\textwidth]{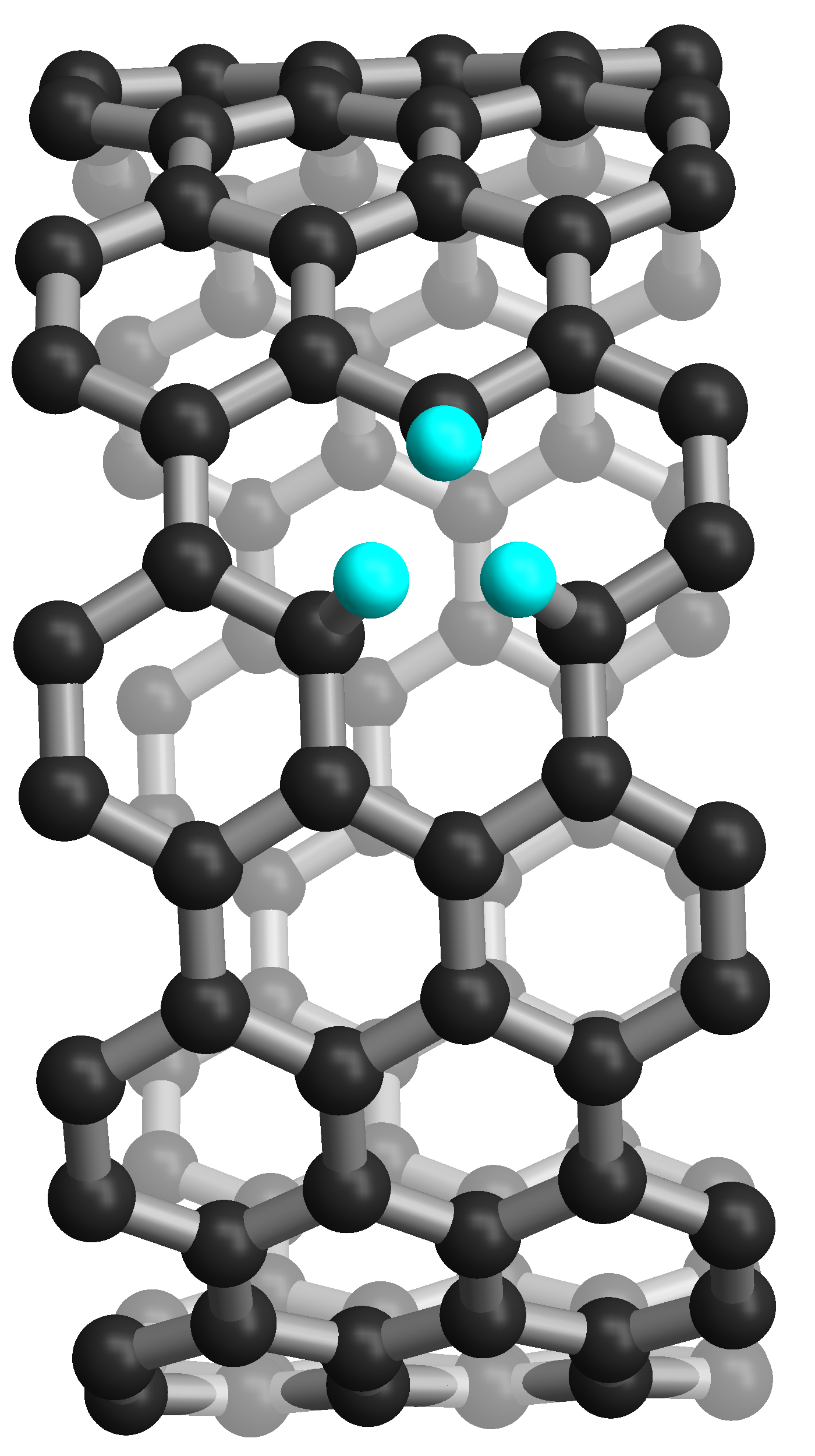}\end{minipage}\hfill
	\begin{minipage}{    \len}\includegraphics[width=\textwidth]{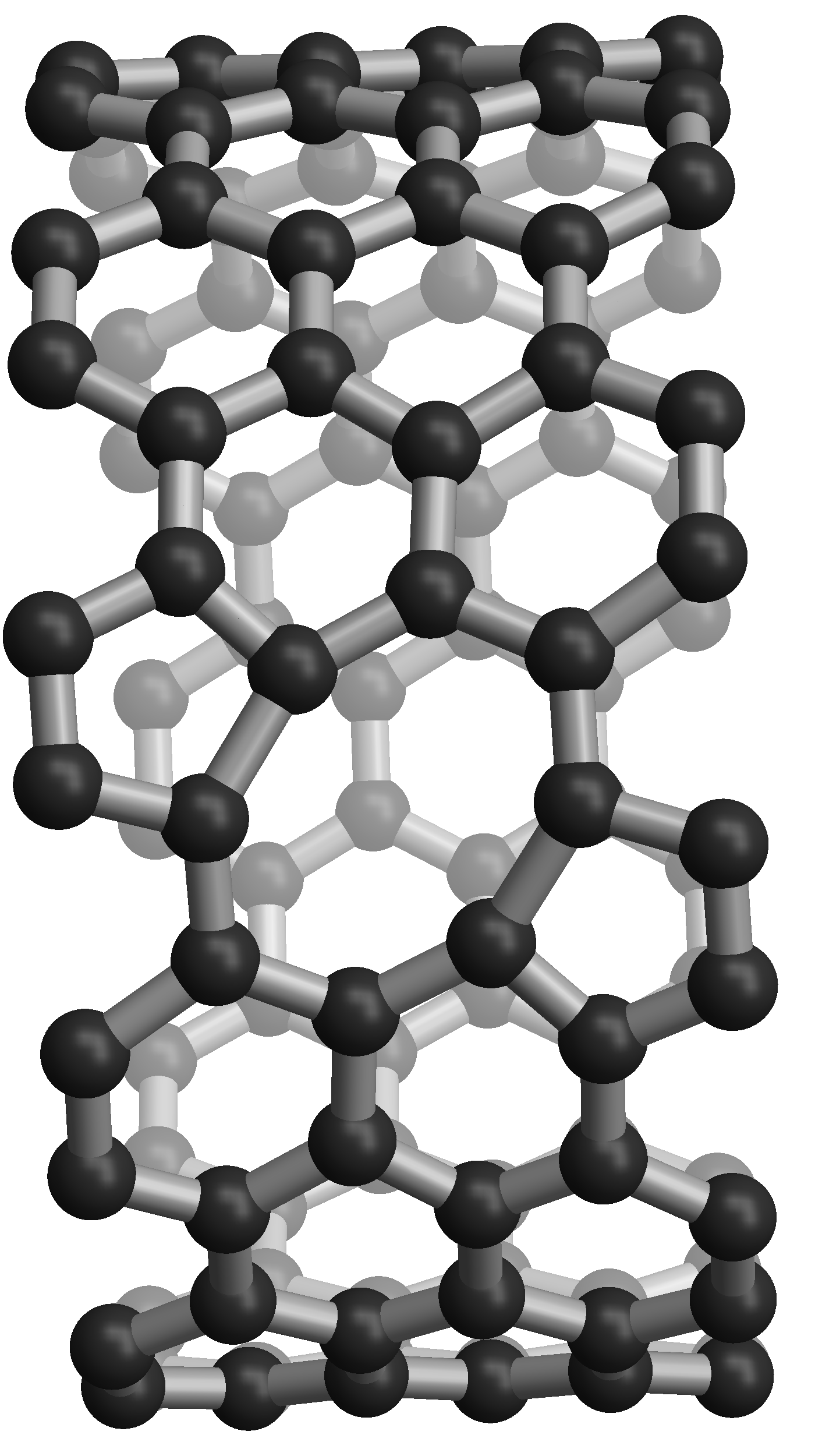}\end{minipage}\hfill
	\begin{minipage}{    \len}\includegraphics[width=\textwidth]{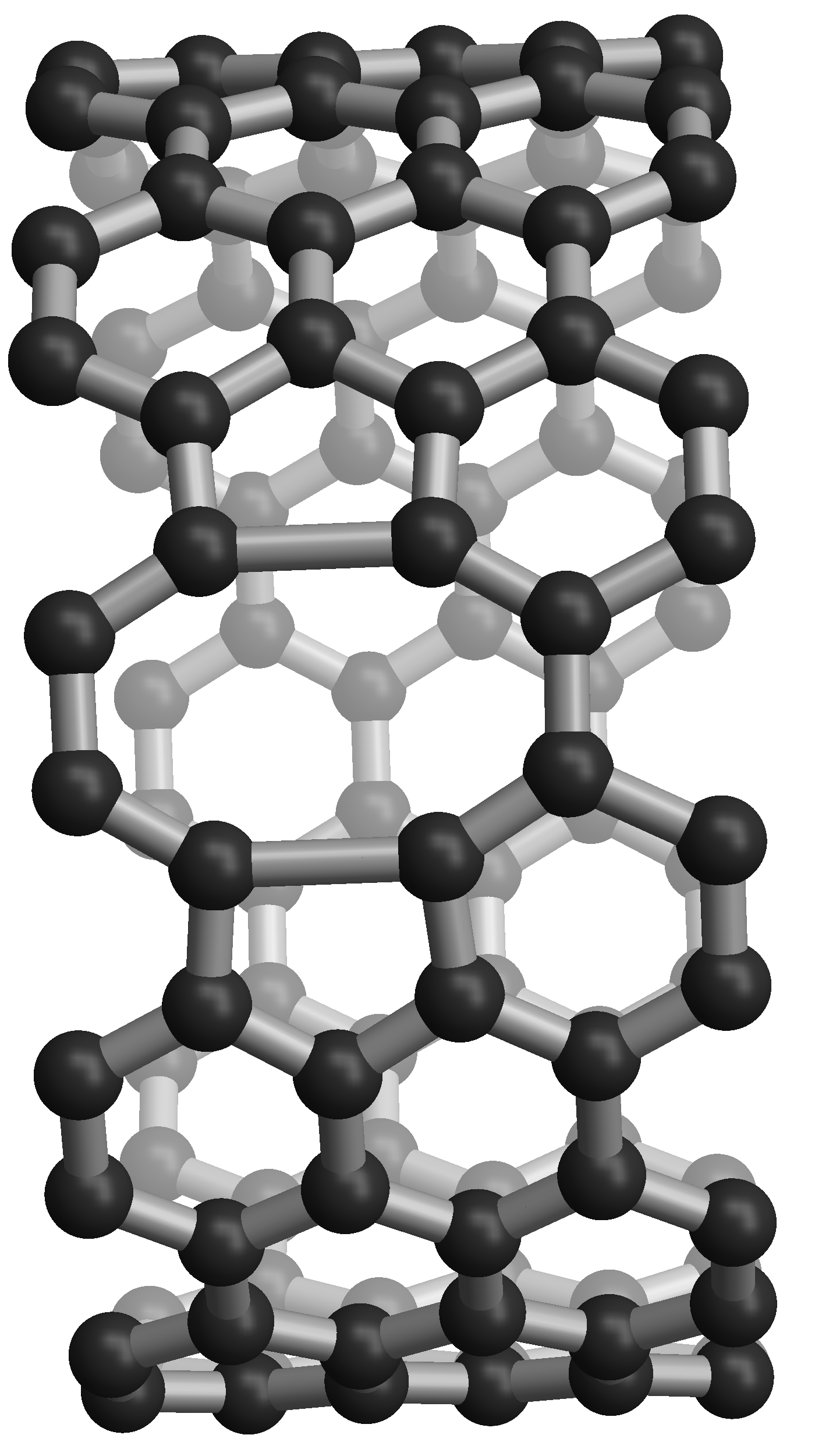}\end{minipage}
	\caption[Geometric structures of the CNT cells]{(Color online.) Geometric structure of the (5,5)-CNT cells and the (10,10)-CNT cells. From left to right: ideal unit cell (UC), unpassivated monovacancy (MV), passivated monovacancy (MV$_\text{3H}$), divacancy in a diagonal orientation (DV$_\text{diag}$) and divacancy in a perpendicular orientation (DV$_\text{perp}$).}
	\label{CMS:fig:CNT}
\end{articlefigure}

For the transport calculation, the electronic structure is described by a density-functional based tight binding model (DFTB)~\cite{PhysRevB.51.12947, IntJQuantumChem.58.185}.
We use the 3ob parameter set, which is a non-orthogonal sp$^3$ basis developed for organic molecules~\cite{JChemTheoryComput.9.338, DFTB}.
The cutoff for the distance-dependent TB hopping energy integrals and the overlap integrals is chosen twice the carbon-carbon distance.
Beyond this distance, the matrix entries are sufficiently small to be neglected.
This cutoff is also favorable, because it leads to not more than third-nearest-neighbor interactions.

We analyze two different types of metallic CNTs, the (5,5)-CNT and the (10,10)-CNT, in combination with three different defect types (see figure \ref{CMS:fig:CNT}): the unpassivated monovacancy (MV), where one carbon atom is removed, the passivated monovacancy, where one carbon atom is removed and the dangling bonds are saturated with hydrogen, and the divacancy (DV), where two neighboring carbon atoms are removed.
The latter can be present in two different orientations: perpendicular to the tube axis (DV$_\text{perp}$) or diagonal (DV$_\text{diag}$).
The length of the MV is equal to the length of the unit cell (UC).
The length of the MV$_\text{3H}$ and the DV$_\text{perp}$ is three times the length of the UC.
For the DV$_\text{diag}$ two cases have to be considered.
The length of the one shown in figure \ref{CMS:fig:CNT} is three times the length of the UC.
Another one, which is mirrored, has four times the length of the UC, because half a UC has to be added at each side to get correct connections to the rest of the CNT.
Because the defect cells (except the MV) are larger than the TB cutoff, they can be further divided into smaller cells to speed up the RGF algorithm.

\begin{articlefigure}[t]
	\includegraphics{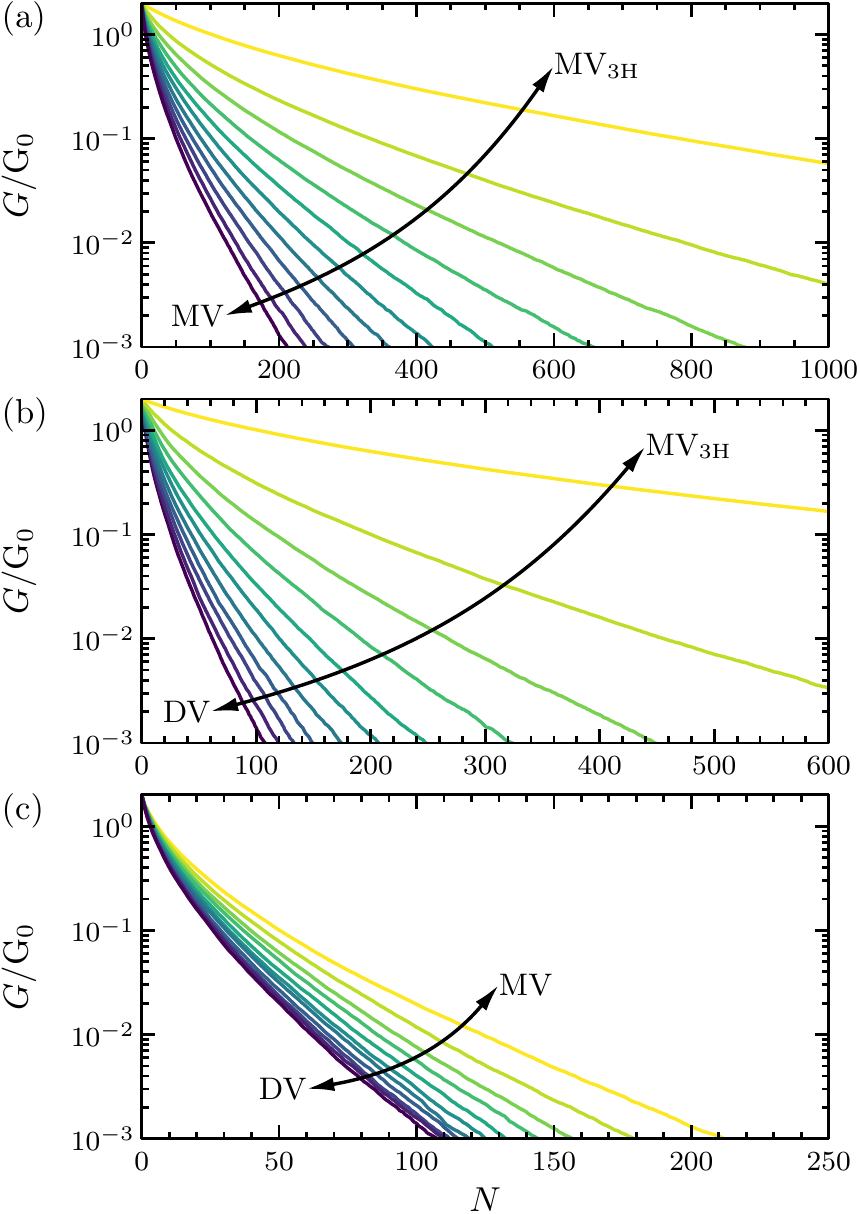}
	\caption[Conductance as a function of the number of defects]{(Color online.) Conductance $G$ as a function of the number of defects $N$ in a (10,10)-CNT at $T=300\,\text{K}$. (a) MV and MV$_\text{3H}$, (b) MV$_\text{3H}$ and DV, (c) DV and MV. Defect fractions $p_\text{MV}$, $p_{\text{MV}_\text{3H}}$, and $p_\text{DV}$ are varied in steps of 0.1 (denoted by color).}
	\label{CMS:fig:G(ND)}
\end{articlefigure}

A realistic structure is obtained by performing a geometry optimization of the ideal unit cell.
Afterwards, the hydrogen atoms and the whole DV defect cell are also optimized.
Finite size effects are taken care of by two additional UCs at each side with the outermost atoms kept fixed at their positions before the optimization.
The DV defect relaxes to one eight-atom ring and two five-atom rings, where the CNT is a bit dented.
The geometry optimization is done with the software Atomistix ToolKit~\cite{ATK.12.8.2, PhysRevB.65.165401}, using density functional theory with the local density approximation of Perdew and Zunger~\cite{PhysRevB.23.5048}, norm-conserving Troullier-Martins pseudopotentials~\cite{PhysRevB.43.1993}, and a SIESTA type double zeta plus double polarization basis set~\cite{JPhysCondMat.14.2745}.

We study CNTs with $N$ realistic defects, whose positions are chosen randomly in an otherwise ideal CNT.
For this, the previously obtained UC and $N$ defect cells are assembled in such a way that the total central region of the CNT has a length of $10N$ UCs.
The defects are randomly distributed in lateral direction as well as in angular direction.
Positions and alignments which cannot be achieved by rotation are achieved by mirroring the atoms at (chiral) lines on the cylinder surface.
In the following, we first consider CNTs with $N$ defects of the same type.
Second, we consider CNTs with $N$ defects of different types $X$, where the defect fractions $p_X$ of the corresponding types are varied.
As the electronic transport properties depend on the exact configuration, i.e. the defect positions, an ensemble of 1000 defective CNTs is considered.
The transmission spectra and conductances are depicted within this work for the ensemble average.

\articlesection{Results and discussion}

\articlesubsection{Conductance}

We calculate the transmission spectrum (\ref{CMS:eqn:TransmissionRGF}) as a function of the number of defects and afterwards the temperature\hyp{}dependent conductance (\ref{CMS:eqn:Conductance}).
In addition to~\cite{NJPhys.16.123026}, we consider mixtures of the defects, shown in figure \ref{CMS:fig:CNT}, by varying the three corresponding defect fractions $p_\text{MV}$, $p_{\text{MV}_\text{3H}}$, and $p_\text{DV}$, with
\begin{articleequation}
	p_\text{MV} + p_{\text{MV}_\text{3H}} + p_\text{DV} = 1 \quad .\label{CMS:eqn:sum_p=1}
\end{articleequation}%

\begin{articlefigure*}[tb]
	\includegraphics{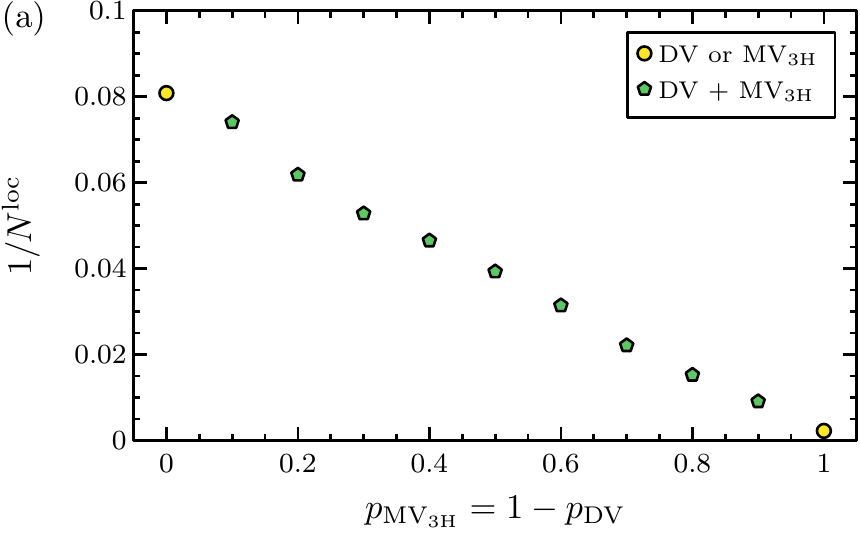}\hfill
	\includegraphics{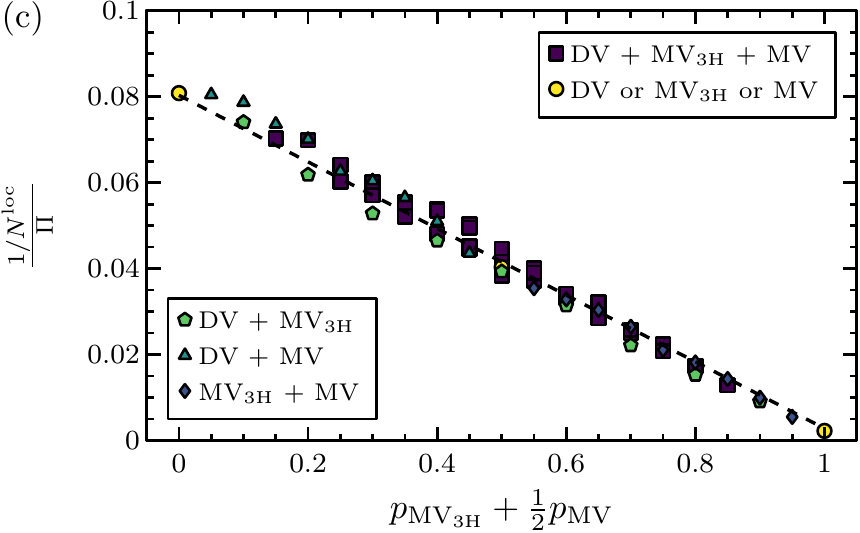}\\[1em]
	\begin{minipage}{\OneColumnWidth}
		\includegraphics{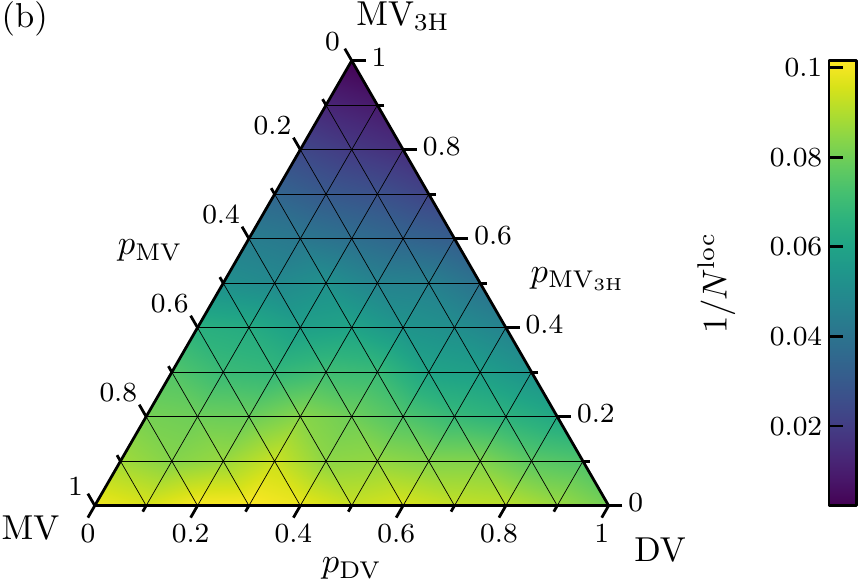}
	\end{minipage}\hfill
	\begin{minipage}{\OneColumnWidth}
		\caption[Localization exponent for the (10,10)-CNT as a function of the defect fractions]{(Color online.) (a) Localization exponent $N^\text{loc}$ for the (10,10)-CNT with a mixture of MV$_\text{3H}$ defects and DV defects at $T=0\,\text{K}$ as a function of the defect fraction $p_{\text{MV}_\text{3H}}$. (b) Localization exponent $N^\text{loc}$ for the (10,10)-CNT with a mixture of MV defects, MV$_\text{3H}$ defects, and DV defects at $T=0\,\text{K}$ as a function of the defect fractions $p_\text{MV}$, $p_{\text{MV}_\text{3H}}$, and $p_\text{DV}$. $N^\text{loc}$ is denoted by color. (c) Same as (b), but in a 2D plot with projected data points (described by $\varPi$) and a corresponding regression (black, dashed). The yellow dots are on the corners of (b), the blueish/greenish triangles, diamonds, and pentagons are on the edges of (b) and the dark blue squares are within the area of (b). The DV+MV$_\text{3H}$ data points (greenish pentagons) are the ones of (a).}\label{CMS:fig:loc(p)}
	\end{minipage}
\end{articlefigure*}

\begin{articlefigure*}
	\includegraphics{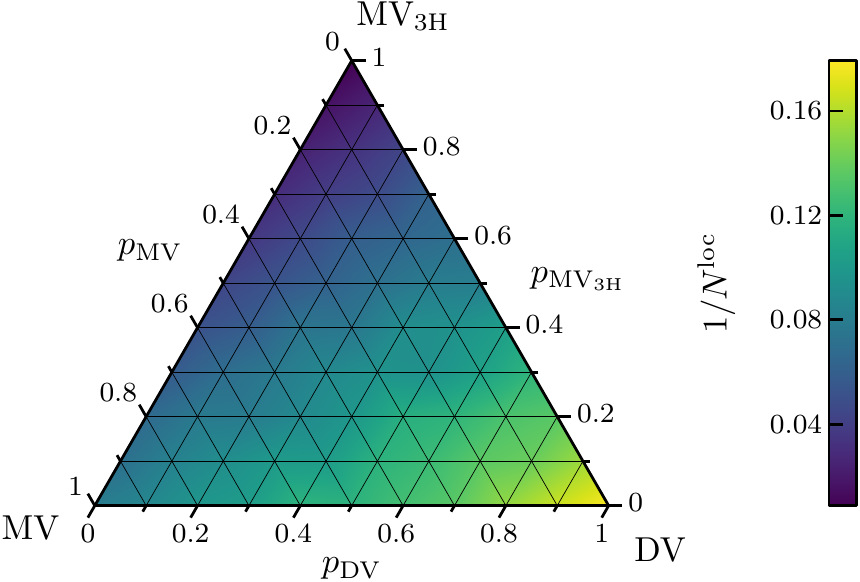}\hfill
	\includegraphics{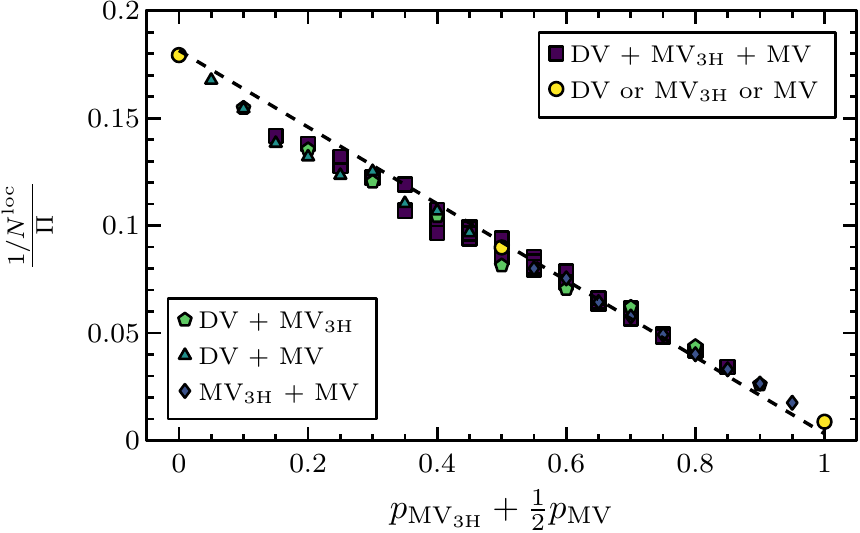}
	\caption[Localization exponent for the (5,5)-CNT as a function of the defect fractions]{(Color online.) Same as figure \ref{CMS:fig:loc(p)}(b) and \ref{CMS:fig:loc(p)}(c), but for the (5,5)-CNT.}\label{CMS:fig:loc(p)_5}
\end{articlefigure*}

Figure \ref{CMS:fig:G(ND)} shows the conductance as a function of the number of defects for (10,10)-CNTs with a mixture of two defects and for different defect fractions at $T=300\,\text{K}$.
At other temperatures, the data look qualitatively equal, but get less smooth for small $T$, especially at $T=0\,\text{K}$.
In the limit of many defects, or in other words, at low conductance, there is an exponential decrease.
This dependence is expected in the strong localization regime, where the states are exponentially localized in space, which leads to exponentially suppressed electron transport with increasing length.
The limit of high conductance (less defects) differs from that, because the states are possibly only weakly localized, omitting the exponential tails.
The conductance in the strong localization limit can be described by
\begin{articleequation}
	G \simeq \text{e}^{-L/\ell^\text{loc}} = \text{e}^{-N/N^\text{loc}} \label{CMS:eqn:Loc}
\end{articleequation}%
with the CNT length $L$, the number of defects $N$, the localization exponent $N^\text{loc}$ and the localization length $\ell^\text{loc} = N^\text{loc}/\rho$, where $\rho$ is the defect density (number of defects per length).
This means that the presence of $N^\text{loc}$ additional defects lowers the conductance $G$ by a factor of $\text{e}$.
In terms of the localization length and a constant defect density, this means that the conductance of a CNT of length $L+\ell^\text{loc}$ is a factor of $\text{e}$ lower compared to the conductance of a CNT of length $L$.

The influence of the different defect fractions is determined by calculating $N^\text{loc}$ via a regression of $G(N)$ and further analysis.
For the regression region, we choose the interval $10^{-3}<G<1.4 \cdot 10^{-1}$.
The results are explained in the next section.

\articlesubsection{Localization exponent}

In the following, we first focus on the low-temperature limit $T=0\,\text{K}$.
Figure \ref{CMS:fig:loc(p)}(a) shows the inverse of the localization exponent as a function of the defect fraction $p_{\text{MV}_\text{3H}}$ for the (10,10)-CNT with MV$_\text{3H}$ and DV defects.
The leftmost data point corresponds to the CNT with only DV defects, the rightmost to the CNT with only MV$_\text{3H}$ defects.
The figure shows a clear linear dependence.
It follows that the overall localization exponent of CNTs with defect mixtures can be written as a weighted harmonic average.
In our case, we have
\begin{articleequation}
	\frac{1}{N^\text{loc}} = \frac{p_{\text{MV}_\text{3H}}}{N_{\text{MV}_\text{3H}}^\text{loc}} + \frac{p_\text{DV}}{N_\text{DV}^\text{loc}} \quad .\label{CMS:eqn:Loc:Mixture:2}
\end{articleequation}%
$N_{\text{MV}_\text{3H}}^\text{loc}$ and $N_\text{DV}^\text{loc}$ are the localization exponents of the CNT with only MV$_\text{3H}$ or DV defects.
The weights are the corresponding defect fractions $p_{\text{MV}_\text{3H}}$ and $p_\text{DV}$.
We emphasize that, instead of performing time-consuming quantum transport methods, (\ref{CMS:eqn:Loc:Mixture:2}) can be used to determine the localization exponents of CNTs with arbitrary defect mixtures if $N_{\text{MV}_\text{3H}}^\text{loc}$ and $N_\text{DV}^\text{loc}$ are known.

Figure \ref{CMS:fig:loc(p)}(b) shows the inverse localization exponent of CNTs with mixtures of all three kinds of defects in a trigonal plot.
The corners correspond to the CNTs with only one defect type, giving $N_\text{MV}^\text{loc}$, $N_{\text{MV}_\text{3H}}^\text{loc}$, and $N_\text{DV}^\text{loc}$.
The edges correspond the CNTs with two of the three defect types.
E.g. considering only the right edge ($p_{\text{MV}_\text{3H}}$) leads to figure \ref{CMS:fig:loc(p)}(a).
The data points within the triangle correspond to proper defect mixtures of all three defects.
The color denotes the inverse localization exponent and shows the same linear dependence for the third defect.
We get the weighted harmonic average
\begin{articleequation}
	\frac{1}{N^\text{loc}} = \frac{p_\text{MV}}{N_\text{MV}^\text{loc}} + \frac{p_{\text{MV}_\text{3H}}}{N_{\text{MV}_\text{3H}}^\text{loc}} + \frac{p_\text{DV}}{N_\text{DV}^\text{loc}} \label{CMS:eqn:Loc:Mixture:3}
\end{articleequation}%
for the overall localization exponent, where $N_\text{MV}^\text{loc}$ is the localization exponent of the CNT with only MV defects.

\begin{articlefigure*}[t]
	\begin{minipage}{\OneColumnWidth}(a) (5,5)-CNT\end{minipage}\hfill
	\begin{minipage}{\OneColumnWidth}(b) (10,10)-CNT\end{minipage}\\[0.5em]
	\includegraphics{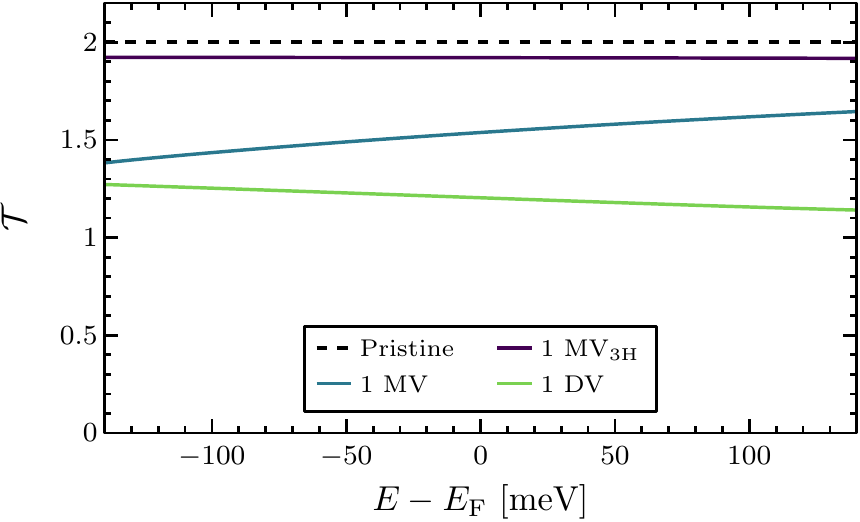}\hfill
	\includegraphics{content/CMS/diagrams/T-E_5.pdf}
	\caption[Transmission spectra]{(Color online.) Transmission functions around the Fermi energy (a) for the (5,5)-CNT and (b) for the (10,10)-CNT.}\label{CMS:fig:temperature:T,f}
\end{articlefigure*}

To get a better idea how well the data of figure \ref{CMS:fig:loc(p)}(b) fit the linear dependence, they are illustrated in another way in figure \ref{CMS:fig:loc(p)}(c).
It shows a tilted side view, where all data points are located as good as possible at one line.
In practice, a two-dimensional linear regression\linebreak $1/N^\text{loc}_\text{reg}(p_\text{MV}$, $p_{\text{MV}_\text{3H}},p_\text{DV})$ of $1/N^\text{loc}$ according to (\ref{CMS:eqn:Loc:Mixture:3}) which fulfills condition (\ref{CMS:eqn:sum_p=1}) was done.
Figure \ref{CMS:fig:loc(p)}(c) now shows a parallel side view onto this regression plane.\footnote{This means the values $1/N^\text{loc}$ of figure \ref{CMS:fig:loc(p)}(b) are (i) divided by the regression $1/N^\text{loc}_\text{reg}(p_\text{MV},p_{\text{MV}_\text{3H}},p_\text{DV})$, giving the relative relations, (ii) projected onto the right edge of figure \ref{CMS:fig:loc(p)}(b), leading to the parameters $p_{\text{MV}_\text{3H}}+\frac{1}{2}p_\text{MV}$ and $p_\text{DV}+\frac{1}{2}p_\text{MV}$ in the two-parameter space, and (iii) rescaled by\linebreak $1/N^\text{loc}_\text{reg}(0,p_{\text{MV}_\text{3H}}+\frac{1}{2}p_\text{MV},p_\text{DV}+\frac{1}{2}p_\text{MV})$, which is the regression on the right edge after the projection.
	The corresponding scaling of $1/N^\text{loc}$ in figure \ref{CMS:fig:loc(p)}(c) is described by\linebreak $\varPi = \frac{1/N_\text{reg}^\text{loc}(p_\text{MV},p_{\text{MV}_\text{3H}},p_\text{DV})}{1/N^\text{loc}_\text{reg}(0,p_{\text{MV}_\text{3H}}+\frac{1}{2}p_\text{MV},p_\text{DV}+\frac{1}{2}p_\text{MV})}$.}

It can be seen that all data fit nearly perfectly with only small deviation, confirming the linear dependence.
The deviations can be explained (i) by the limited ensemble size, which might be not large enough to obtain smooth curves $G(N)$ from the average and (ii) by the fact that the region $N$ we used for the exponential regression of $G(N)$ is not completely in the strong localization regime.

As discussed for the (10,10)-CNT so far, figure \ref{CMS:fig:loc(p)_5} shows the respective calculations for the (5,5)-CNT.
An identical behavior (\ref{CMS:eqn:Loc:Mixture:3}) is found, but with different impacts of different defect types.
Especially the localization exponent of the MV is bigger than the one of the DV, whereas for the (10,10)-CNT the relation is the other way round.
However, this indicates that (\ref{CMS:eqn:Loc:Mixture:3}) can be assumed independent of the CNT type.

Equation (\ref{CMS:eqn:Loc:Mixture:3}) can also be extended to other defect types.
An easy explanation can be given by considering isolated defects, which contribute to the exponential conductance reduction in the strongly localized regime without a disturbance by other defects.
Each additional defect of type $X$ leads to $G(M+1)=G(M)\,\text{exp}(-1/N^\text{loc}_X)$.
In total, when considering $N_X$ additional defects of types~$X$, relation (\ref{CMS:eqn:Loc:Mixture:3}) follows from (\ref{CMS:eqn:Loc}).
As a consequence, the total localization exponent $N^\text{loc}$ of defect mixtures can be calculated by the localization exponents $N_X^\text{loc}$ of the single defect types. 
In table \ref{CMS:tab:loc}, the localization exponents at $T=0\,\text{K}$ and $T=300\,\text{K}$ of the pure defect types, extracted from (\ref{CMS:eqn:Loc}) and figure \ref{CMS:fig:G(ND)}, are listed.

\begin{articletable}[!b]
	\begin{tabular}{l|l|l|l}
		Defect         & $T$ [K] & (5,5)-CNT & (10,10)-CNT \\
		\hline
		MV             &   0     &  11       &  10 \\ 
		               & 300     &  13       &  35 \\ 
		\hline
		MV$_\text{3H}$ &   0     & 110       & 460 \\ 
		               & 300     & 110       & 450 \\ 
		\hline
		DV             &   0     &  5.2      &  15 \\ 
		               & 300     &  6.0      &  17    
	\end{tabular}
	\caption[Localization exponents of (5,5)- and (10,10)-CNTs with defects of one type]{Localization exponents of (5,5)- and (10,10)-CNTs with defect of one type (MV, MV$_\text{3H}$, or DV).}\label{CMS:tab:loc}
\end{articletable}

\begin{articlefigure*}[!b]
	\begin{minipage}{0.49\textwidth}(a) (5,5)-CNT\end{minipage}\hfill
	\begin{minipage}{0.49\textwidth}(b) (10,10)-CNT\end{minipage}\\[0.5em]
	\includegraphics{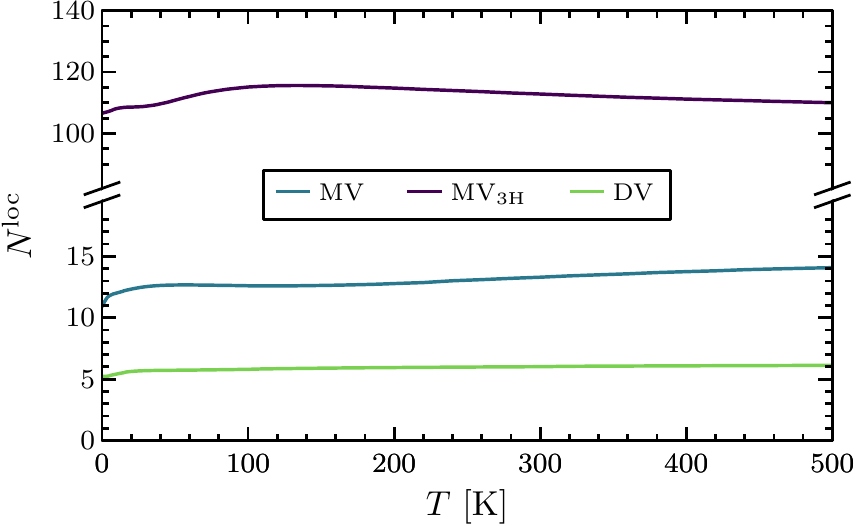}\hfill
	\includegraphics{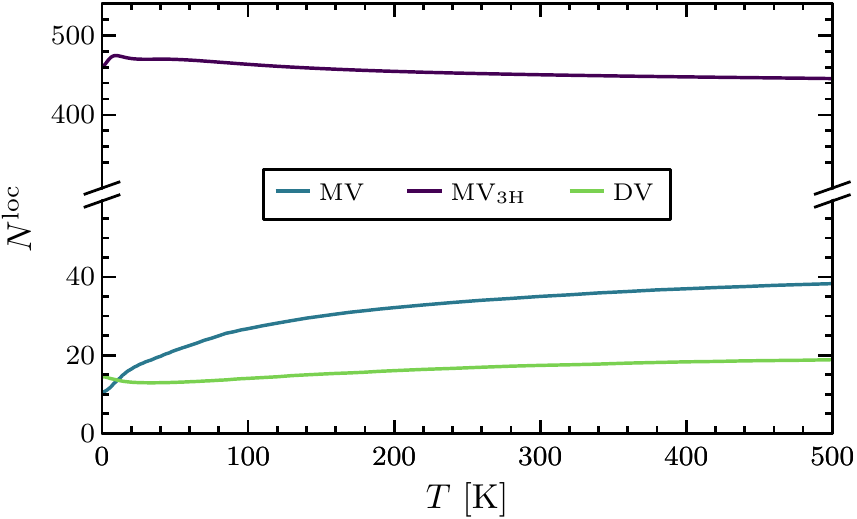}
	\caption[Temperature dependence of the localization exponent]{(Color online.) Temperature dependence of the localization exponents $N^\text{loc}_\text{MV}$, $N^\text{loc}_{\text{MV}_\text{3H}}$, and $N^\text{loc}_\text{DV}$ (a) for the (5,5)-CNT and (b) for the (10,10)-CNT.}\label{CMS:fig:temperature:loc}
\end{articlefigure*}

A comparison with experiments is difficult, as the specific type of the defects and their fractions are not known exactly.
Also a systematic variation of the defect fractions has not yet been presented.
This makes the direct experimental validation of (\ref{CMS:eqn:Loc:Mixture:3}) hardly possible.
Thus, we can only relate experimental and theoretical results, estimate the missing structural information, and discuss whether its order of magnitude physically makes sense or not.
In \cite{NatureMaterials.4.534}, a localization length of\linebreak $\ell^\text{loc}=(230\pm120)\,\text{nm}$ for a (10,10)-CNT or one with similar diameter has been extracted from measurements.
In the present study dimensionless localization exponents $N^\text{loc}$ are calculated.
Both can be related using (\ref{CMS:eqn:Loc}), yielding $\ell^\text{loc}=dN^\text{loc}$, where $d$ is the average defect distance.
We assume only DV defects and neglect MV$_\text{3H}$ defects because of their much smaller impact on the conductance, as also done in \cite{NatureMaterials.4.534}.
At $T=0\,\text{K}$ we get $N^\text{loc}=15$ for this specific case.
Taking both, the localization length and the localization exponent, the average defect distance is $d=(15\pm 8)\,\text{nm}$, which is equivalent to the length of\linebreak $36\pm 19$ UCs or a defect probability per atom of\linebreak $(0.7\pm 0.4)\permil$.
This is within a experimentally observable region and agrees with $d=(14\ldots 75)\,\text{nm}$ reported in \cite{NatureMaterials.4.534}.

\articlesubsection{Influence of temperature}

\begin{articlefigure*}[t]
	\begin{minipage}{0.49\textwidth}(a) (5,5)-CNT\end{minipage}\hfill
	\begin{minipage}{0.49\textwidth}(b) (10,10)-CNT\end{minipage}\\[0.5em]
	\includegraphics{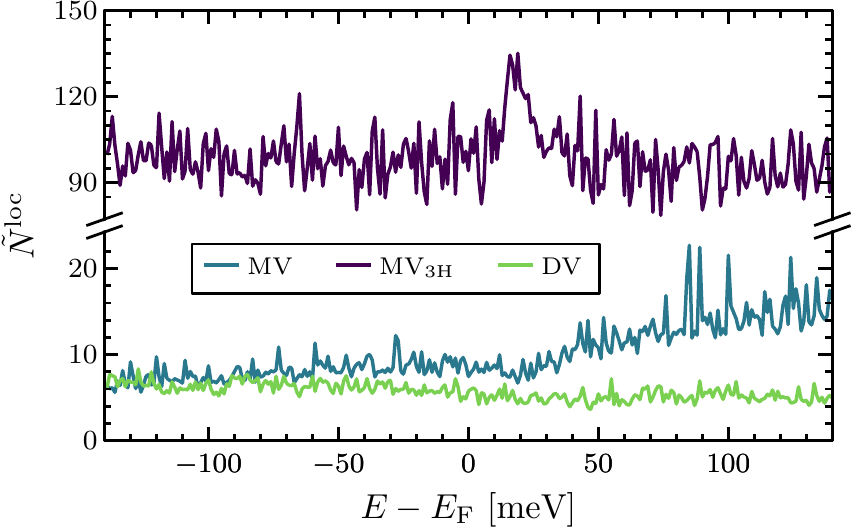}\hfill
	\includegraphics{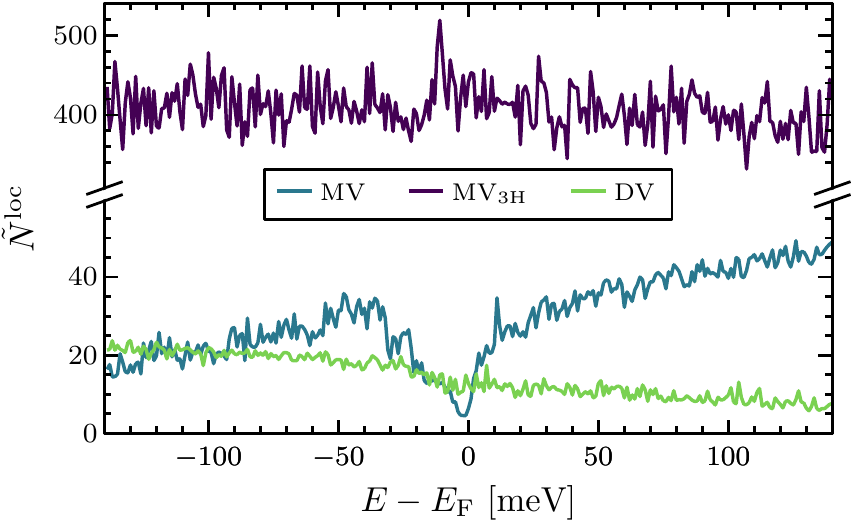}
	\caption[Energy dependence of the localization exponent]{(Color online.) Energy dependence of the localization exponents $N^\text{loc}_\text{MV}$, $N^\text{loc}_{\text{MV}_\text{3H}}$, and $N^\text{loc}_\text{DV}$ (a) for the (5,5)-CNT and (b) for the (10,10)-CNT.}\label{CMS:fig:temperature:loc(E)}
\end{articlefigure*}

In general, the localization exponent depends on the temperature.
This is taken into account via the Fermi distribution in the Landauer formula (\ref{CMS:eqn:Conductance}).
The previously performed calculations have also been extended to different temperatures $T\neq 0\,\text{K}$.
The resulting $N^\text{loc}$ for mixed defects yield data very similar to what is shown in figure \ref{CMS:fig:loc(p)} (and are thus not additionally depicted).
The qualitative dependencies are the same with just other absolute values.
Thus, relation (\ref{CMS:eqn:Loc:Mixture:3}) keeps valid and we only have to discuss the temperature dependence of the localization exponents for CNTs with a single defect type.

Figure \ref{CMS:fig:temperature:T,f} shows the transmission function $\mathcal{T}(E)$ for a CNT with one of the three different defects MV, MV$_\text{3H}$, and DV.
Only the transmission of the (10,10)-CNT with a MV defect (figure \ref{CMS:fig:temperature:T,f}(b)) has pronounced features in the plotted energy range.
The other spectra are nearly linear or only slightly curved.
The conductance for a given transmission function $\mathcal{T}(E)$ is temperature-dependent if $\mathcal{T}(E)$ is not linear around the Fermi energy, especially if there are defect-induced features, which is only the case for the (10,10)-CNT with a MV defects.
Otherwise it is constant due to the symmetry of $\text{d}f/\text{d}E$ in (\ref{CMS:eqn:Conductance}).
The origin why the transmission of the MV$_\text{3H}$ defect is that much closer to the ideal case than the others could be the maintained undistorted benzoidal structure in contrast to the dangling bonds of the MV defect and the local reconstruction of the DV defect.

Figure \ref{CMS:fig:temperature:loc} shows the temperature-dependent localization exponents of the (5,5)- and the (10,10)-CNT with either MV, MV$_\text{3H}$, or DV defects.
Only $N^\text{loc}_\text{MV}$ of the (10,10)-CNT shows a significant temperature dependence, where $N^\text{loc}_\text{MV}$ increases with increasing temperature by a factor of 4.
This is caused by the defect-induced feature in the transmission function $\mathcal{T}(E)$, which is located directly at the Fermi energy.
In this case, at low temperature, the deep valley $\mathcal{T}(E=E_\text{F})$ is determining the conductance~$G$.
Because of the strong reduction of $G$ compared to the pristine CNT, the localization exponent is small.
For increasing temperature, the higher transmission at energies around the Fermi energy leads to a higher conductance.
In this case, the reduction of $G$ compared to the pristine CNT is less strong and thus, the localization exponent is higher.
In contrast to the (10,10)-CNT with MV defects, only slight temperature influences can be seen for $N^\text{loc}_\text{DV}$ (both CNT types), $N^\text{loc}_{\text{MV}_\text{3H}}$ (both CNT types), and $N^\text{loc}_\text{MV}$ (the (5,5)-CNT).
The relative influence is of the order of 10\%.

In summary, a strong energy dependence of the transmission around the Fermi energy, which leads to a temperature dependence of the conductance, is necessary for a significant temperature dependence of the localization exponent.
This can be seen in figure \ref{CMS:fig:temperature:loc} for the (10,10)-CNT with MV defects.

To confirm the previous explanations, figure \ref{CMS:fig:temperature:loc(E)} shows the energy-resolved localization exponent of the transmission, which is calculated for each of the energies separately via a regression (with respect to $N$) of
\begin{articleequation}
	\mathcal{T}(N,E) \simeq \text{e}^{-N/\tilde{N}^\text{loc}(E)} \quad .
\end{articleequation}%
The numerical fluctuations are high due to the finite CNT ensemble and the resulting fluctuations in the transmission function for CNTs with many defects, but the qualitative trends can be seen.
$\tilde{N}^\text{loc}_\text{DV}$ (both CNT types), $\tilde{N}^\text{loc}_{\text{MV}_\text{3H}}$ (both CNT types), and $\tilde{N}^\text{loc}_\text{MV}$ (only the (5,5)-CNT) are quite flat or even constant.
Only $\tilde{N}^\text{loc}_\text{MV}$ of the (10,10)-CNT has a significant feature directly at the Fermi energy, which is in agreement with the feature in the transmission spectrum.
In this case, at low temperature, where only the Fermi energy is relevant, the localization exponent is small (dip in figure \ref{CMS:fig:temperature:loc(E)}(b) for the MV).
At higher temperature, where the other energies contribute more and more, the localization exponent increases with increasing temperature, as shown in figure \ref{CMS:fig:temperature:loc}(b).

Finally, we note that MV defects are predicted to be unstable, because dangling bonds have a strong tendency to get saturated.
Consequently, the influence of temperature on the localization exponent of the vacancy defects in reality is small.

\articlesubsection{Conductance estimation}

Equation (\ref{CMS:eqn:Loc}) describes the scaling behavior of the conductance with respect to the number of defects in the strong localization regime.
We showed, that the dependence of the total localization exponent of defect mixtures on the localization exponent of single defect types is described by (\ref{CMS:eqn:Loc:Mixture:3}).
Both can be combined in the single expression
\begin{articleequation}
	G\left(L,\rho,\left\{p_X\right\}\right) = \tilde{g}\,\text{exp}\left(-L\rho\sum_X\frac{p_X}{N^\text{loc}_X}\right)\label{CMS:eqn:Gestimation}
\end{articleequation}%
to estimate and predict the conductance of metallic CNTs with arbitrary defect configurations. $L$ is the length of the CNT, $\rho$ is the total defect density (number of defects per length), and $p_X$ is the defect fraction of defect type $X$.
The localization exponents $N^\text{loc}_X$ depend on temperature, CNT type and defect type and have been determined previously.
Once they have been calculated for each defect type $X$, they can be used for arbitrary mixtures of such defects, as we have shown for (5,5)- and (10,10)-CNTs with different vacancies.
This can be adopted to other CNTs and defect types.

In \cite{NJPhys.16.123026}, we showed that the diameter dependence of $N^\text{loc}$ for armchair CNTs is linear.
This can be used to further estimate the conductance of CNTs with large diameter.

A last thing to mention is that the regression also provides the prefactor of (\ref{CMS:eqn:Gestimation}).
Our results show that it does not depend on the defect type or the defect\linebreak fractions besides a large unsystematic variance.\linebreak
We get $\tilde{g}=(0.4\ldots 0.6)\text{G}_0$ for the (5,5)-CNT and\linebreak $\tilde{g}=(0.3\ldots 0.5)\text{G}_0$ for the (10,10)-CNT.
Furthermore,\linebreak (\ref{CMS:eqn:Gestimation}) does not describe the region of small defect numbers but the limit of large disorder, which is approximately reached for $G<0.1\text{G}_0$.

\articlesection{Summary and conclusions}

We investigated the electronic transport properties of mesoscopic metallic CNTs with mixtures of realistic defects in the strong localization regime.

Based on the analysis of the scaling behavior of the CNT conductance with respect to the number of defects of different types, we showed that the localization exponent can be written as the weighted harmonic average of the localization exponents of CNTs with identical defects of one type.
As a consequence, the localization exponent of CNTs with defect mixtures can be estimated and predicted, omitting time-consuming calculations of large CNT ensembles with statistical distributions of different defect types.

The effect of temperature on the localization exponent is present, but altogether small for the metallic CNTs studied in this work.
Considering semiconducting CNTs, this effect should be much larger, because the bandgap results in a highly non-linear transmission function $\mathcal{T}(E)$ near the Fermi energy (around the band edges).
This will be subject of future work.

Finally, we brought all results together and provide a conductance model for defective metallic CNTs within the strong localization regime.
This is an important step towards the description of defective CNTs in the mesoscopic range.
If the CNTs are larger in diameter and more defect types are involved, the computations become increasingly unfeasible.
So, for a technological utilization of CNTs, our results can be used to estimate the electronic transport properties of defective metallic CNTs with dimensions that cannot be treated with the common Green's-function-based quantum transport calculations.

\articlesection*{Acknowledgement}

This work is funded by the European Union (ERDF) and the Free State of Saxony via the ESF project\linebreak 100231947 (Young Investigators Group Computer Simulations for Materials Design - CoSiMa).
\begin{center}
	\includegraphics[width=0.35\textwidth]{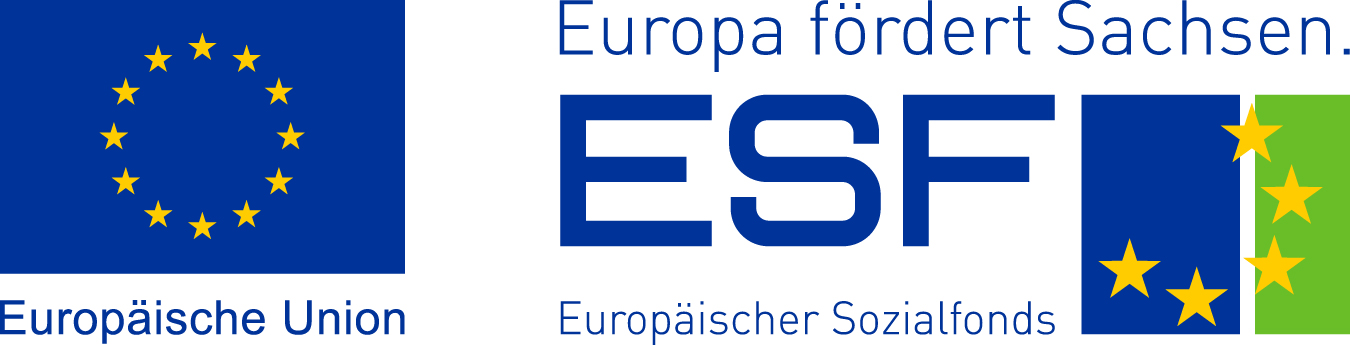}
\end{center}

\end{document}